\documentclass[11pt]{article}
\usepackage{color}
\usepackage{amssymb}   
\usepackage{amsthm}    
\usepackage{amsmath}   
\usepackage{stmaryrd}  
\usepackage{titletoc}  
\usepackage{mathrsfs}  
\usepackage{graphicx}
\usepackage{multicol}
\usepackage{multirow}
\usepackage{subcaption}
\usepackage{caption}
\usepackage{float}
\usepackage{hyperref}
\usepackage{cleveref}
\usepackage{xcolor}
\usepackage{hyperref}
\usepackage[english]{babel}
\addto\extrasenglish{
  
}
\addto\extrasenglish{
  
}
\addto\extrasenglish{
  
}
\usepackage{comment}

\hypersetup{
  colorlinks   = true, 
  urlcolor     = red, 
  linkcolor    = blue, 
  citecolor   = blue 
}
\usepackage[round]{natbib}   
\newlength{\defbaselineskip}
\newcommand{\setlinespacing}[1]%
           {\setlength{\baselineskip}{#1 \defbaselineskip}}

\makeatletter\@addtoreset{equation}{section} \makeatother
\DeclareMathOperator*{\argmin}{arg\,min}
 \allowdisplaybreaks
\begin{document}
\title{Broken Adaptive Ridge Method for Variable Selection in Generalized Partly Linear Models with Application to the Coronary Artery Disease Data}

\author{Christian Chan$^1$  \and Xiaotian Dai$^2$ \and Thierry Chekouo$^3$ \and Quan Long$^4$ \and Xuewen Lu$^1$ \\
\small{$^1$ Department of Mathematics and Statistics, University of Calgary} \\
\small{$^2$ Department of Mathematics, Illinois State University} \\
\small{$^3$ Division of Biostatistics, School of Public Health, University of Minnesota} \\ 
\small{$^4$ Department of Biochemistry and Molecular Biology, University of Calgary } \\
}

\maketitle

\begin{abstract}
Motivated by the CATHGEN data, we develop a new statistical learning method for simultaneous variable selection and parameter estimation under the context of generalized partly linear models for data with high-dimensional covariates. The method is referred to as the broken adaptive ridge (BAR) estimator, which is an approximation of the $L_0$-penalized regression by iteratively performing reweighted squared $L_2$-penalized regression. The generalized partly linear model extends the generalized linear model by including a non-parametric component to construct a flexible model for modeling various types of covariate effects. We employ the Bernstein polynomials as the sieve space to approximate the non-parametric functions so that our method can be implemented easily using the existing R packages. Extensive simulation studies suggest that the proposed method performs better than other commonly used penalty-based variable selection methods. We apply the method to the CATHGEN data with a binary response from a coronary artery disease study, which motivated our research, and obtained new findings in both high-dimensional genetic and low-dimensional non-genetic covariates. 
\end{abstract}


\section{Introduction}
In the era of high technology and supercomputing power, the combination of lower financial costs and greater accessibility to DNA sequencing technology has contributed to the rapid rise in omics research \citep{mardis2011decade}. The high-throughput DNA sequencing equipment produces high-dimensional data, which motivates researchers to identify the genetic variations in the genome that are relevant to the phenotype. In our research, we develop new statistical learning methods to further decode acquired data and help scientists to find relevant genetic covariates.  Variable selection is one of these statistical learning methods and is an important task when building a statistical model. Given a large number of explanatory variables in a particular study, we want to select the variables that are relevant to the response variable. One way to do this is by using the best subset selection, which is based on the $L_0$-regularization. The best subset selection method directly penalizes the cardinality of a model subject to an information criterion, like the AIC \citep{akaike1974new} and BIC \citep{schwarz1978estimating}. There are several disadvantages to the $L_0$-regularization method, the most important of them is the computational complexity at scales of $2^p$, where $p$ is the dimension of the covariates, thus making it computationally expensive for even a moderately large number of covariates. Additionally, \citet{breiman1996heuristics} also showed that the $L_0$-regularization is unstable in terms of variable selection. Penalty-based variable selection methods were introduced to solve the computation inefficiency of $L_0$-regularization. The significance of penalty-based variable selection is the reformulation of the sparse estimation problem into a continuous and nonconvex or convex optimization problem with a fewer number of candidate models. Such methods include the Least Absolute Shrinkage and Selection Operator (LASSO) \citep{tibshirani1996regression}, Smoothly Clipped Absolute Deviation (SCAD) \citep{fan2001variable}, the Elastic-Net penalty \citep{zou2005regularization}, the Adaptive LASSO \citep{zou2006adaptive} and the Minimax Concave Penalty (MCP) \citep{zhang2010nearly}. 

Recently, the Broken Adaptive Ridge (BAR) regression method has been introduced as an approximation to the $L_0$-regularization for variable selection. The BAR regression can be summarized as an iteratively reweighted squared $L_2$-penalized regression, where the estimators of the BAR method are taken at the limit of the algorithm. \cite{liu2016efficient} first considered the implementation of the BAR method under generalized linear models (GLM). Since then, many papers have investigated the BAR method for different models and data types, including the Cox PH model with large-scale right-censored survival data \citep{kawaguchi2020surrogate}, the linear model with uncensored data \citep{dai2018broken}, the additive hazards model with recurrent event data \citep{zhao2018variable}, the Cox PH model with interval-censored data \citep{zhao2019simultaneous}, the partly linear Cox PH model with right-censored data \citep{wu2020variable}, and the accelerated failure time model with right-censored data \citep{sun2022broken}, among others. Most recently, \cite{mahmoudi2022penalized} incorporated the BAR method for semi-competing risks data under the illness-death model. Previous work \citep{dai2018broken,zhao2018variable,kawaguchi2020surrogate} have also proved that the BAR method possesses two desired large-sample properties: consistency for variable selection and asymptotic normality, which are called the oracle properties in the literature. 

Motivated by the CATHGEN data detailed below, our goal in this research is to extend the BAR method to select important variables in generalized partly linear models (GPLMs) with a large number of genetic covariates, in the presence of some low-dimensional non-genetic covariates. Particularly, we apply the proposed method to select important single nucleotide polymorphisms (SNPs) in a logistic partly linear model in the presence of both categorical and continuous low dimensional non-genetic covariates, which belongs to a family of generalized partly linear models. In the CATHerization GENetics (CATHGEN) study, the primary objective was to assess the association of multiple genetic markers with cardiovascular disease phenotypes. The study, conducted by Duke University Medical Centre, collected peripheral blood samples from consenting patients between 2001 until 2012. The follow-up period of the recruited patients was between 2004 and 2014. Aside from the high-dimensional genetic data, low-dimensional baseline clinical and demographical variables were also measured when patients were first recruited to the study. The data can be downloaded from U.S. National Institute of Health dbGaP data accession number phs000704.v1.p1. We will use the proposed method to analyze the data to identify important SNPs and the associated genes relevant to coronary artery disease (CAD). 

The contributions of our work can be summarized from three main aspects. {\bf First}, we develop a new statistical learning method for simultaneous variable selection and estimation under the context of generalized partly linear models using the BAR method. GPLMs extend GLMs by adding a non-parametric component to it to allow for flexible modeling of linear and non-linear covariate effects. Our method extends the work by \cite{li2021scalable} which incorporates the BAR method under GLM with sparse high-dimensional and massive sample size data for variable selection. Since the low-dimensional covariates in our work contain both categorical and continuous variables, our method treats them separately, only the continuous variables are considered to possess potential non-linear effects. {\bf Second}, we focus specifically on the logistic partly linear regression model, as motivated by the presence of the binary response variable (CAD vs. no CAD) and various types of covariates in the CATHGEN data. We apply our proposed method to the CATHGEN data to identify relevant genetic markers (i.e., SNPs) in high dimensions that contribute to developing CAD. We are also interested in the estimation of the low-dimensional relevant non-genetic covariate effects, which can handle both linear and non-linear covariate effects. {\bf Third}, our method can be easily implemented using existing R packages developed for GLM, since we are able to use Bernstein polynomials to construct a linear sieve space for estimating the non-parametric functions of the low-dimensional covariates so that the resultant model form mimics a GLM and the existing GLM R packages can be used for estimation and variable selection. We make our code available at \url{https://github.com/chrischan94/GPLM-BAR}.

The rest of this article is organized as follows. In \autoref{methodology}, we give a comprehensive introduction to GPLMs, and a detailed explanation of our proposed method and its algorithm. In \autoref{sec:simstudies}, we present the results of our extensive simulation studies, where we compare our method to a few common variable selection methods. In \autoref{rda}, we present the results of the real data analysis of the CATHGEN study and the biological interpretation of the results. Finally, in \autoref{sec:diss}, we conclude our findings in this article and discuss possible future directions for research. The description of the involved algorithm and more simulation results are relegated to the Appendix. 

\section{Models and Methods} \label{methodology}

\subsection{Generalized Partly Linear Models} 
Working under the framework of GLM, consider a random sample $\textbf{v}_i = (\textbf{x}_i^\top,y_i)^\top,i=1,\ldots,n$, where 
$\textbf{y} = \{y_1,\ldots,y_n \}^\top$ makes a $n \times 1$ response vector  and  matrix $\textbf{X}=\{\textbf{x}_1,\ldots,\textbf{x}_n\}^\top$ makes a $n \times p$ design. The observations $\textbf{v}_i = (\textbf{x}_i^\top,y_i)^\top,i=1,\ldots,n$ are mutually independent. 
The distribution of $y_i$ conditional on $\textbf{x}_i$ is from the exponential family with the following density, 
\begin{equation}\label{expfamily}
f_\textbf{y}(y_i;\theta_i, \phi) =\exp\left\{ \frac{y_i\theta_i - b(\theta_i)}{a(\phi)}+ c(y_i,\phi) \right\},
\end{equation}
where $a(\cdot)$, $b(\cdot)$ and $c(\cdot,\cdot)$ are known specific functions, $b(\cdot)$ is assumed to be twice differentiable, $\theta_i$ is the canonical parameter and $\phi$ denotes the dispersion parameter. Model \eqref{expfamily} indicates that $\mathbb{E}(y_i | \textbf{x}_i) =\mu_i = b'(\theta_i)$ and $\mathbb{V}(y_i | \textbf{x}_i) = b''(\theta_i) a(\phi)$.  Through a link function $g(\mu_i) = \boldsymbol{\beta}^\top \textbf{x}_i$, the canonical parameter $\theta_i$ is connected to $\textbf{x}_i$ by a linear combination of the coefficient parameter vector $\boldsymbol{\beta} = \{\beta_1,\ldots,\beta_p \}^\top$. When $g(\mu_i) = \theta_i$, it is called the canonical link function. Commonly used canonical link functions in GLMs include the identity function for linear regression, logit link function for logistic regression and the log function for Poisson regression. Given our observed data, the likelihood function of $\boldsymbol{\beta}$ for GLMs is
$$
\mathcal{L}_n(\boldsymbol{\beta}; \textbf{v}_i) = \prod^n_{i=1} f_\textbf{y}(y_i;\theta_i, \phi) = \prod^n_{i=1} \exp\left\{ \frac{y_i\theta_i - b(\theta_i)}{a(\phi)}+ c(y_i,\phi) \right\},
$$
and the log-likelihood is
$$
\ell_n(\boldsymbol{\beta}) = \log \mathcal{L}_n(\boldsymbol{\beta};\textbf{v}_i) = \sum^n_{i=1} \log f_\textbf{y}(y_i;\theta_i, \phi).
$$
GLM assumes a linear relationship between the independent variables $\{\textbf{x}_1,\ldots,\textbf{x}_n\}^\top$ and the canonical link function. If this assumption is violated for a subset of variables that may have a non-linear relationship with the response, an alternative model form is desirable. Motivated by the CATHGEN data that contains both genetic and non-genetic variables, the covariates can be broadly grouped into three distinct sets: a $n \times p$ design matrix $\textbf{X}=\{\textbf{x}_1,\ldots,\textbf{x}_n\}^\top$, a $n \times q_w$ design matrix $\textbf{W} = \{\textbf{w}_1,\ldots,\textbf{w}_n \}^\top$, and a $n \times q_z$ design matrix $\textbf{Z} = \{\textbf{z}_1,\ldots,\textbf{z}_n \}^\top$. The design matrix $\textbf{X}$ contains the high-dimensional genetic covariates, $\textbf{W}$ contains the low-dimensional and non-genetic categorical covariates, and $\textbf{Z}$ contains the low-dimensional and non-genetic continuous covariates. Then we define the generalized partly linear model as follows, which extends GLM by adding a non-parametric component in the linear predictor,
$$
g\left\{\mathbb{E}(y_i | \textbf{x}_i, \textbf{w}_i, \textbf{z}_i)\right\} = \boldsymbol{\beta}^\top \textbf{x}_i + \boldsymbol{\alpha}^\top \textbf{w}_i + \Psi(\textbf{z}_i),
$$
where  $\boldsymbol{\beta} = \{\beta_1, \ldots, \beta_p \}^\top$, $\boldsymbol{\alpha}=\{\alpha_1, \ldots ,\alpha_{q_w} \}^\top$,  $\Psi(\textbf{z}_i) = \sum^{q_z}_{j=1} \psi_j(z_{ij})$, $\psi_j(\cdot)$'s are unknown smooth functions, which model possible non-linear effects as shown in the analysis of the CATHGEN data. A special case of GPLM is the logistic partly linear model. Let $\pi(y_i|\textbf{x}_i,\textbf{w}_i,\textbf{z}_i) = P(y_i = 1|\textbf{x}_i,\textbf{w}_i,\textbf{z}_i)$, then the logistic partly linear model has the model equation given by
$$
\log{\frac{\pi(y_i|\textbf{x}_i,\textbf{w}_i,\textbf{z}_i)}{1-\pi(y_i|\textbf{x}_i,\textbf{w}_i,\textbf{z}_i)}} = \boldsymbol{\beta}^\top \textbf{x}_i + \boldsymbol{\alpha}^\top \textbf{w}_i + \Psi(\textbf{z}_i).
$$
For the observations 
$\{\textbf{u}_i,i=1,\ldots,n\}= \{\{y_i, \textbf{x}_i, \textbf{w}_i, \textbf{z}_i\}, i=1,\ldots,n\}$, the likelihood function of the logistic partly linear model can be constructed as
$$
\mathcal{L}_n(\boldsymbol{\alpha}, \boldsymbol{\beta}, \boldsymbol{\psi}) = \prod^n_{i=1} \pi(y_i|\textbf{x}_i,\textbf{w}_i,\textbf{z}_i)^{y_i} \left\{1-\pi(y_i|\textbf{x}_i,\textbf{w}_i,\textbf{z}_i ) \right\}^{1-y_i},
$$
where $\boldsymbol{\psi}=\{\psi_1(\cdot),\ldots,\psi_{q_z}(\cdot) \}^\top$.
From the likelihood function, the log-likelihood can be easily derived as follows
\begin{equation}\label{PLLRMlog}
\begin{split}
\ell_n(\boldsymbol{\alpha},\boldsymbol{\beta},\boldsymbol{\psi}) & = \sum^n_{i=1}  \big[y_i \left(\boldsymbol{\beta}^\top \textbf{x}_i + \boldsymbol{\alpha}^\top \textbf{w}_i + \Psi(\textbf{z}_i)\right) \\ 
& - \log\left(1+ \exp\left\{\boldsymbol{\beta}^\top \textbf{x}_i + \boldsymbol{\alpha}^\top \textbf{w}_i + \Psi(\textbf{z}_i) \right\}\right)\big].
\end{split}
\end{equation}
Direct estimation of $\boldsymbol{\vartheta} = (\boldsymbol{\alpha}, \boldsymbol{\beta}, \boldsymbol{\psi})$ in \eqref{PLLRMlog} will not be possible because of the presence of the unknown functions $\Psi(\textbf{z}_i)$ which are infinitely dimensional. Hence, approximating the non-parametric part is needed. As the unknown functions $\Psi(\textbf{z}_i)$ are infinitely dimensional, we propose to construct a sieve space to linearize them. To apply the sieve method, we employ the Bernstein polynomials to approximate $\Psi(\textbf{z}_i)$, then, the Bernstein polynomials approximation reduces the infinitely dimensional space to a finitely dimensional space. Let $\boldsymbol{\Theta}$ denote the parameter space of $\boldsymbol{\vartheta}$ where
$$
\boldsymbol{\Theta} = \left\{ \boldsymbol{\vartheta}=\left(\boldsymbol{\alpha},\boldsymbol{\beta},\psi_1,\ldots,\psi_{q_z}\right) \in \mathcal{A} \otimes \mathcal{M}_1 \otimes \cdots \otimes \mathcal{M}_{q_z}\right\},
$$
where 
$$
\mathcal{A} = \left\{\left(\boldsymbol{\alpha}, \boldsymbol{\beta}\right) \in \mathbb{R}^{q_w} \times \mathbb{R}^p, \left\lVert \boldsymbol{\alpha}\right\rVert + \left\lVert \boldsymbol{\beta}\right\rVert \leq M \right\},
$$
$M$ is a positive constant, and $\mathcal{M}_j$ denotes the collection of all bounded and continuous functions over the range of the observed $\textbf{z}_j$ for $j=1,\ldots,q_z$. Subsequently, the sieve space is defined as
$$
\boldsymbol{\Theta}_n = \left\{\boldsymbol{\vartheta}_n = \left(\boldsymbol{\alpha},\boldsymbol{\beta},\psi_{1n},\ldots,\psi_{q_{z}n}\right) \in \mathcal{A} \otimes \mathcal{M}_{1n} \otimes \cdots \otimes \mathcal{M}_{q_{z}n}\right\},
$$
where
\begin{equation}\label{approx}
\mathcal{M}_{jn} = \left\{ \psi_{jn} (z_{ij}) = \sum^{m_j}_{k=0} \gamma_{jk} B_{jk} (z_{ij}, m_j, c_j, u_j) : \sum_{0 \leq k \leq m_j} |\gamma_{jk}| \leq M_{jn} \right\},
\end{equation}
for $j=1,\ldots,q_z$ and $z_{ij} \in [c_j, u_j]$, $c_j<u_j$. The Bernstein basis polynomial of $m_j$ degree, denoted by $B_{jk} (z_{ij}, m_j, c_j, u_j)$ in \eqref{approx}, has the equation
$$
B_{jk}(z_{ij}, m_j, c_j, u_j) = \begin{pmatrix} m_j \\ k \end{pmatrix} \Bigg( \frac{z_{ij} - c_j}{u_j - c_j} \Bigg)^k \Bigg(1 - \frac{z_{ij} - c_j}{u_j - c_j} \Bigg)^{m_j - k}, \quad k = 0,\ldots,m_j.
$$
Therefore, the sieve log-likelihood of the logistic partly linear model using the Bernstein polynomial to approximate $\Psi(\textbf{z}_i)$ is
\begin{equation}\label{loglikeLPLM}
\begin{split}
\ell_n(\boldsymbol{\alpha}, \boldsymbol{\beta}, \boldsymbol{\gamma}) = & \sum^n_{i=1} \bigg\{ y_i  \Big(\boldsymbol{\beta}^\top \textbf{x}_i + \boldsymbol{\alpha}^\top \textbf{w}_i + \sum^{q_z}_{j=1} \sum^{m_j}_{k=0} \gamma_{jk} B_{jk}(z_{ij},m_j,c_j,u_j) \Big) \\
& -\log\Big(1+ \exp\big\{ \boldsymbol{\beta}^\top \textbf{x}_i + \boldsymbol{\alpha}^\top \textbf{w}_i + \sum^{q_z}_{j=1} \sum^{m_j}_{k=0} \gamma_{jk} B_{jk}(z_{ij},m_j,c_j,u_j)\big\} \Big) \bigg\},
\end{split}
\end{equation}
where $\boldsymbol{\gamma} = \{ \gamma_{10},\ldots,\gamma_{1m_{1}},\ldots,\gamma_{q_{z}0},\ldots,\gamma_{q_{z}m_{q_z}} \}^\top $.

\subsection{Simultaneous Estimation and Variable Selection using GPLM-BAR method}
To conduct simultaneous estimation and variable selection in GPLMs, we propose the GPLM-BAR method, which is an iterative method. Following the BAR method by \citet{li2021scalable} for GLM, starting from an initial value vector  computed from the following 
the ridge regression,
\begin{equation}\label{GPLMBARinit}
\left(\widehat{\boldsymbol{\alpha}}^{(0)},\widehat{\boldsymbol{\beta}}^{(0)},\widehat{\boldsymbol{\gamma}}^{(0)}\right) = \argmin_{\boldsymbol{\alpha},\boldsymbol{\beta},\boldsymbol{\gamma}} \left\{-2\ell_n(\boldsymbol{\alpha}, \boldsymbol{\beta},\boldsymbol{\gamma}) + \xi_n \sum^p_{j=1} \beta^2_j \right\}.
\end{equation}
For $s \geq 1$, the estimator is iteratively updated by a reweighted squared $L_2$-penalized regression
\begin{equation}\label{GPLMBARup}
\left(\widehat{\boldsymbol{\alpha}}^{(s)},\widehat{\boldsymbol{\beta}}^{(s)},\widehat{\boldsymbol{\gamma}}^{(s)} \right) = \argmin_{\boldsymbol{\alpha}, \boldsymbol{\beta},\boldsymbol{\gamma}} \left\{-2 \ell_n(\boldsymbol{\alpha}, \boldsymbol{\beta},\boldsymbol{\gamma}) + \lambda_n \sum^p_{j=1} \frac{\beta^2_j}{(\widehat{\beta}^{(s-1)}_j)^2} \right\},
\end{equation}
where $\xi_n$ and $\lambda_n$ are non-negative tuning parameters. The updated step in \eqref{GPLMBARup} is continued until a pre-specified convergence criterion is reached, where the estimators are taken at the limit as
$$
\widehat{\boldsymbol{\beta}} = \lim_{s \rightarrow \infty} \widehat{\boldsymbol{\beta}}^{(s)}, \quad \widehat{\boldsymbol{\alpha}} =  \lim_{s \rightarrow \infty} \widehat{\boldsymbol{\alpha}}^{(s)}, \quad \widehat{\boldsymbol{\gamma}} =  \lim_{s \rightarrow \infty} \widehat{\boldsymbol{\gamma}}^{(s)}.
$$
The implementation of the proposed method indicates the variable selection is done only on high-dimensional covariates $\textbf{x}_i$, $i=1,\ldots,n$, since the penalty is imposed on $\boldsymbol{\beta}$ only. In addition, the proposed variable selection method can be applied in a similar fashion to  the  Poisson  partly linear regression and the partly linear model for counts and continuous responses, respectively, since they are in the family of GPLMs.  

\subsubsection{A note on choosing the tuning parameters}
\begin{figure}
\centering
\includegraphics[height = 7.75cm, width = 11.2cm]{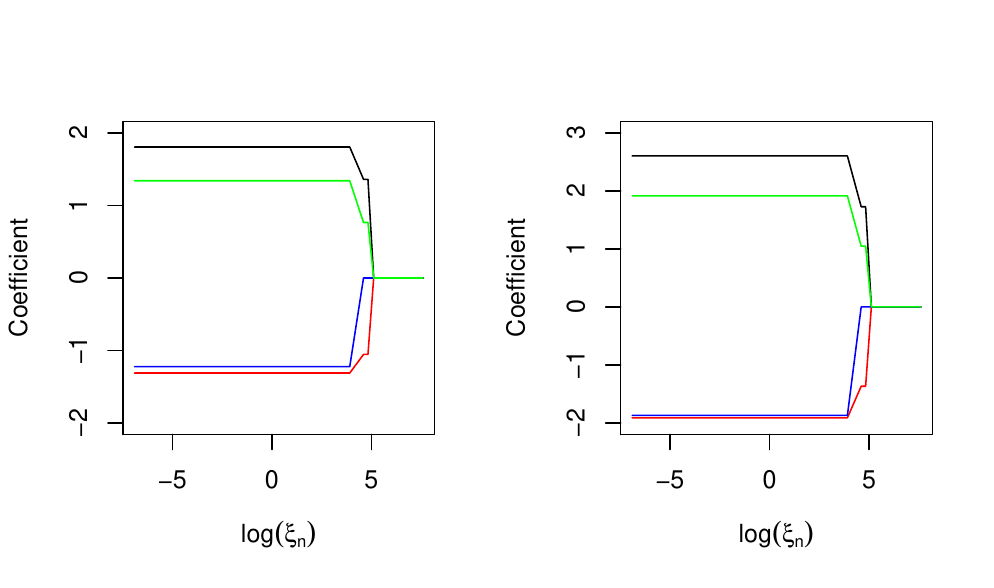}
\caption{Path plots for the logistic partly linear regression with varying $\xi_n$ for a random sample of size $n = 200$ and $p=10$, using both the BIC (left panel) and AIC (right panel) penalties.} \label{path}
\end{figure}

Choosing the optimal values of tuning parameters is crucial for penalty-based variable selection methods, as it greatly affects the variable selection accuracy. In the absence of an external validation set, common methods to find the optimal values of the tuning parameters include the $k$-fold cross-validation (CV) method. The optimal tuning parameter value is the one that minimizes a criterion. Typically, this is the mean squared error for continuous outcome, or deviance for binary outcome. However, doing this only adds to the computational complexity, and it is not ideal for larger datasets. In the GPLM-BAR algorithm, we have two tuning parameters: $\xi_n$ and $\lambda_n$. Unless the value of $\xi_n$ chosen is large, it is empirically shown that the value chosen is inconsequential on the estimation of $\boldsymbol{\beta}$, as seen in Figure \ref{path}. Hence, $\xi_n$ is set to a relatively small value. 
For $\lambda_n$ in the Cox-BAR regression, it has been argued by \citet{kawaguchi2020surrogate}  that it can be fixed. One example is fixing $\lambda_n = \log(n)$, which corresponds the BIC penalty. Another example is to fix $\lambda_n = 2$, which corresponds to the AIC penalty. In our method,  both the AIC and BIC penalties are considered.

\subsubsection{Computational aspects for GPLM-BAR}
Except under the linear model, numerical approximation methods such as the Newton-Raphson algorithm are integrated into the implementation of the BAR penalty for simultaneous variable selection and estimation. When both the number of covariates and sample size are small, calculating the partial gradient vector and Hessian matrix at each iteration of the BAR algorithm is computationally feasible. However, when both the number of covariates and sample size becomes moderately big, numerical approximation becomes not scalable because of the high computational costs and the numerical instability. Alternative optimization techniques for parameter estimation under large-scale regularization and regression problems \citep{zhang2001text,azoury2001relative} have been developed. The algorithm by \cite{zhang2001text} called column relaxation of logistic loss (CLG) can be classified as a cyclic coordinate descent algorithm.

The R package \verb|BrokenAdaptiveRidge| \citep{kawaguchi2020surrogate} was created to implement BAR regression for GLM and the Cox model, which are linear models. Since we have reparameterized our GPLM into a form of GLM in \eqref{loglikeLPLM}, we are able to directly use the package to conduct variable selection and estimation under the context of GPLM. This package uses the R package \verb|Cyclops| \citep{suchard2013massive} for efficient implementation of the iterative method as described in \cite{kawaguchi2020surrogate}. To do this, first we create a data frame \verb|B| for Bernstein polynomial basis functions based on the low-dimensional covariates \verb|Z|, each column in \verb|B(Z)| represents one basis function. Let \verb|X| represent the design matrix for high-dimensional genetic covariates, and \verb|W| for low-dimensional non-genetic covariates, \verb|Y| represents the $(0,1)$ binary response vector. Finally, we make a combined data frame \verb|D|. Then, for example, the GPLM-BAR estimates selected by AIC can be computed from the following R code
\begin{verbatim}
D <- data.matrix(cbind(X,W,B(Z))
penAIC <- createBarPrior(penalty = 2, 
                         exclude = c(1, ((ncol(X)+2):(ncol(D)+1)), 
                         initialRidgeVariance = 1) 
       #penalty=2 indicates AIC, and log(n) indicates BIC;
cyD <- createCyclopsData(Y ~ D, modelType = "lr") 
       #lr indicates logistic regression
BARfit <- fitCyclopsModel(cyD, prior = penAIC)
       #estimates of all of the coefficients
\end{verbatim}
The computation in the package is done by the cyclic coordinate descent algorithm. We describe this algorithm for the GPLM-BAR regression in the Appendix. 

\section{Simulation Studies} \label{sec:simstudies}
In this section, we present the results of a comprehensive simulation study in three scenarios to demonstrate the effectiveness of our proposed method. The first and second scenarios assess the performance under strong signals and weak signals, respectively, under the setting of the logistic partly linear model. The third scenario uses the selected model in the real data analysis section for the CATHGEN data as a basis to simulate data, then assesses the performance of our method, and the fourth scenario shows the performance under the setting of the Poisson partly linear model. 

\subsection*{Scenario 1: Strong signals in the logistic partly linear model}
In this scenario, let $q_z=5$ and $q_w=4$, and the number of non-zero elements $q=5$ in the true parameter $p$-vector $\boldsymbol{\beta}_0$,  for various values of $p$. We generate the $n\times p$ design matrix $\textbf{X}$ from a multivariate normal distribution with mean zero and variance-covariance matrix $\Sigma_\textbf{X}$, where the $(i,j)^{th}$ entry of it is $\rho^{|i-j|}$. We fix $\rho=0.25$. We first consider large effects, i.e., large values of $\boldsymbol{\beta}$, where the true value of $\boldsymbol{\beta}$ is $\boldsymbol{\beta}_0= \{ 1,-1,0,\ldots,0,-1,0.75,0.75 \}^\top$. We also generate a $n \times q_w$ design matrix $\textbf{W}$ from independent Bernoulli distributions, with the same probability of success $\pi=0.5$. And, the true value of $\boldsymbol{\alpha}$ is $\boldsymbol{\alpha}_0= \{1,-0.5,-0.5,0.75,-1\}^\top$. Independently from $\textbf{X}$ and $\textbf{W}$, we also generate a $n \times q_z$ design matrix $\textbf{Z}$, where we draw $\textbf{z}_1$ from the uniform distribution over (1,5), $\textbf{z}_2$ and $\textbf{z}_3$  independently from the standard uniform distribution, and $\textbf{z}_4$ from the uniform distribution over $(-3,1)$. By setting the non-linear functions to be $\psi_1(z_{i1}) = 0.1(z_{i1}-3)^2$, $\psi_2(z_{i2}) = 0.2(\cos(2\pi z_{i2})+1)$, $\psi_3(z_{i3}) = 0.2\sin(2\pi z_{i3})$, and $\psi_4 (z_{i4}) = 0.2(z_{i4}+1)^3$, respectively, we generate $y_i$ from the Bernoulli distribution with probability $\pi_i$, where $\pi_i = 1/(1+\exp\{-\boldsymbol{\beta}_0^\top \textbf{x}_i - \boldsymbol{\alpha}_0^\top \textbf{w}_i - \sum^4_{j=1} \psi_j(z_{ij})\})$. The chosen non-linear functions have two common properties: 1) they are symmetric at the midpoint of the interval of their domains, 2) The values of the functions are zero at the midpoint for the purpose of identifiability. We consider two different sample sizes $n=600$ and $800$, and two different numbers of high-dimensional covariates $p=300$ and $450$. Each combination is replicated 200 times. The number of basis functions for all non-linear functions is set at $m_j+1=3+1=4$, since more than four basis functions only add to the computational complexity while only marginally improving the approximation of $\psi_j(\cdot)$, conversely having fewer than four basis functions will not approximate $\psi_j(\cdot)$ well. 

In the simulation studies, we compare our method against the methods of LASSO and Adaptive LASSO. We use the R package \verb|splines2| \citep{splines2-paper} to generate the Bernstein polynomials. The LASSO and Adaptive LASSO methods are implemented using the R package \verb|glmnet| \citep{glmnet2010,simon2011regularization}.  To evaluate the estimation accuracy, we compute the median mean squared error (MMSE), where the mean squared error has the equation $(\hat{\boldsymbol{\beta}} - \boldsymbol{\beta}_0)^\top \Sigma_\textbf{X} (\hat{\boldsymbol{\beta}} - \boldsymbol{\beta}_0)$. For the GPLM-BAR method, we fix $\lambda_n$ to two values, $\lambda_n=2$ and $\lambda_n = \log(n)$, which corresponds to the AIC and BIC penalties respectively. Since the value of $\xi_n$ was  shown to have an inconsequential effect on estimation, we set $\xi_n = 1$. For the other methods, we use $10$-fold CV method to select the optimal value. To evaluate the selection accuracy, we compute the average number of true positives (TP), average number of false positives (FP), total misclassification rate (MC), frequency of true model (TM) selected, and the average estimated size of the model (MS), where $\text{MS} = \text{TP} + \text{FP}$. 
\begin{table}
\centering
\caption{Variable selection and estimation results over 200 replications for Scenario 1. Standard deviations of the MMSE are in parentheses.} \label{SS}
\begin{tabular}{l|llllll}
\hline
 Method & MMSE & TP & FP & MS & MC & TM \\
\hline
\multicolumn{7}{c}{$n=600,p=300$} \\
\hline
BAR(AIC) & 0.17(0.10) & 5 & 1.31 & 6.31 & 1.31 & 24$\%$ \\
BAR(BIC) & 0.28(0.27) & 4.74 & 0 & 4.74 & 0.26 & 74$\%$ \\
LASSO & 1.03(0.22) & 5 & 3.48 & 8.48 & 3.48 & 16$\%$  \\
ALASSO & 0.46(0.24) & 5 & 2.68 & 7.68 & 2.68 & 51$\%$  \\
Oracle & 0.08(0.08) & 5 & 0 & 5 & 0 & 100$\%$ \\
\hline
\multicolumn{7}{c}{$n=800,p=300$} \\
\hline
BAR(AIC) & 0.11(0.07) & 5 & 1.21 & 6.21 & 1.21 & 28$\%$ \\
BAR(BIC) & 0.17(0.12) & 4.97 & 0 & 4.97 & 0.03 & 98$\%$ \\
LASSO & 0.67(0.17) & 5 & 8.10 & 13.10 & 8.10 & 1$\%$   \\
ALASSO & 0.21(0.11) & 5 & 4.10 & 9.10 & 4.10 & 32$\%$  \\
Oracle & 0.05(0.06) & 5 & 0 & 5 & 0 & 100$\%$ \\
\hline
\multicolumn{7}{c}{$n=600,p=450$} \\
\hline
BAR(AIC) & 0.20(0.14) & 5 & 1.83 & 6.83 & 1.83 & 17$\%$  \\
BAR(BIC) & 0.32(0.31) & 4.70 & 0 & 4.70 & 0.30 & 74$\%$ \\
LASSO & 0.89(0.22) & 5 & 10.47 & 15.47 & 10.47 & 2$\%$  \\
ALASSO & 0.33(0.16) & 5 & 10.35 & 15.35 & 10.35 & 11$\%$  \\
Oracle & 0.08(0.07) & 5 & 0 & 5 & 0 & 100$\%$ \\
\hline
\multicolumn{7}{c}{$n=800,p=450$} \\
\hline
BAR(AIC) & 0.14(0.09) & 5 & 1.77 & 6.77 & 1.77 & 16$\%$  \\
BAR(BIC) & 0.16(0.13) & 4.98 & 0 & 4.98 & 0.02 & 98$\%$ \\
LASSO& 0.70(0.18) & 5 & 9.80 & 14.80 & 9.80 & 1$\%$  \\
ALASSO & 0.22(0.15) & 5 & 7.62 & 12.62 & 7.62 & 19$\%$  \\ 
Oracle & 0.06(0.06) & 5 & 0 & 5 & 0 & 100$\%$ \\
\hline
\end{tabular}
\end{table}
From Table \ref{SS}, one can observe that the GPLM-BAR method performs better than LASSO and Adaptive LASSO by most measures of accuracy. Although the average number of TP is not the best particularly with the BIC penalty, the average number of FP is far lower as compared to the other methods, resulting in the total misclassification rate being the lowest. The GPLM-BAR method also produces the sparsest model. It is interesting to observe a trade-off between the AIC and BIC penalties, where estimation accuracy is better with the AIC penalty, contrasting the better variable selection results with the BIC penalty. This is explained by the larger tuning parameter in the BIC penalty, which shrinks the relatively smaller signals in $\boldsymbol{\beta}$ to zero, thus causing a larger estimation bias. We also report the estimation results of $\boldsymbol{\alpha}$ in Table \ref{alphaS1} in the Appendix, where the estimates of our method is the best.

\begin{figure}
\centering
\centerline{\includegraphics[height=11cm, width=14cm]{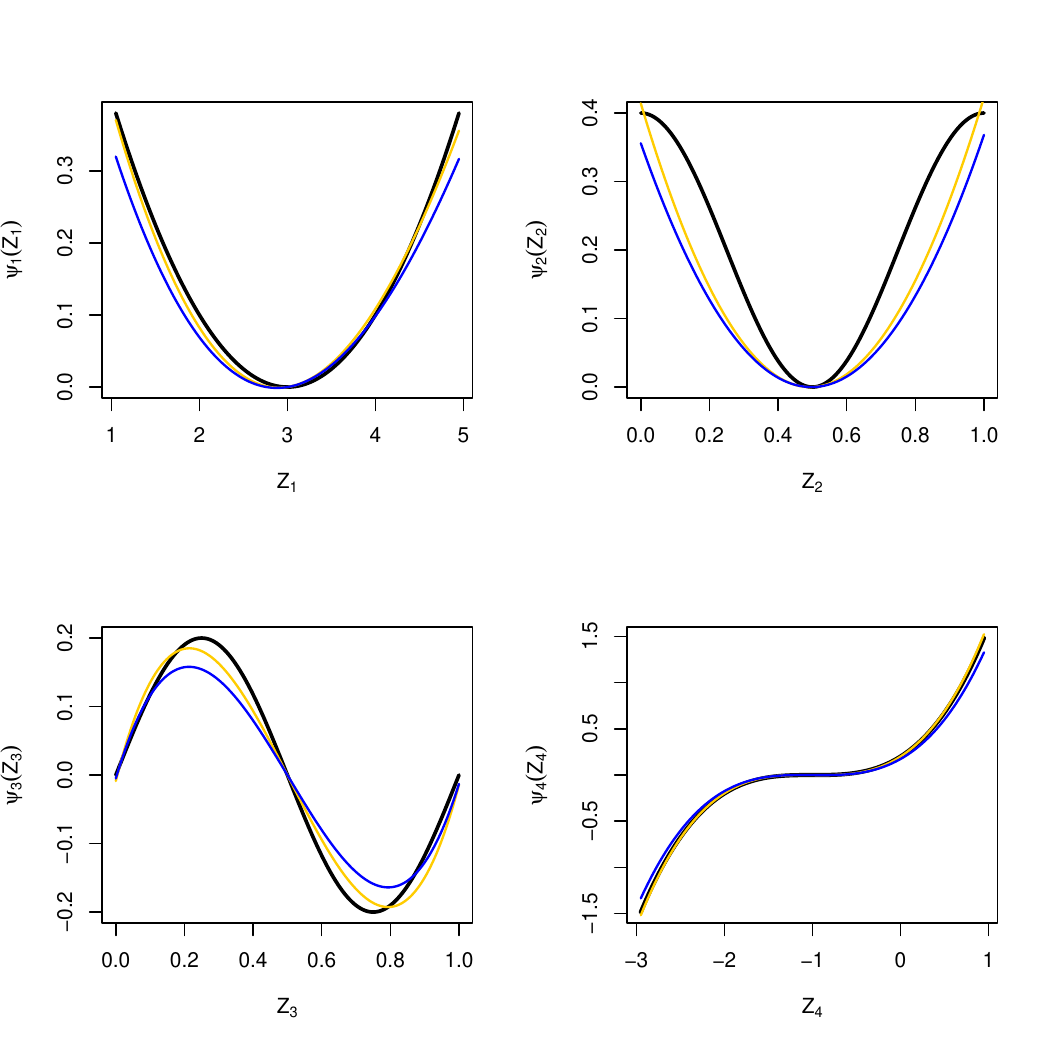}}
\caption{Estimated nonlinear covariate effects $\psi_j(\cdot), j=1,2,3,4$, for Scenario 1, $p=300$ and $n=600$.} \label{nonlinES}
\end{figure}

We also are interested in the estimation of non-linear covariate effects $\psi_j(\cdot)$ using the GPLM-BAR method. The estimated curves are shown in Figure \ref{nonlinES}, which compares the averaged estimates of each of the four non-linear functions to the true function. Two observations are made. First, the Bernstein polynomial using three basis functions to approximate each $\psi_j(\cdot)$ is satisfactory, where the general shape of each function is captured well. Second, different $\lambda_n$ tuning  methods give slightly different estimates of $\psi_j(\cdot)$, as the BAR method with the BIC penalty (blue curve) produces more biases than the AIC penalty (yellow curve). The GPLM-BAR also performs well for the other three combinations of $n$ and $p$ (Appendix Figures \ref{ESTS1p300n800},\ref{ESTS1p450n600} and \ref{ESTS1p450n800}).

\subsection*{Scenario 2: Strong and weak signals in the logistic partly linear model}
We also perform another scenario where a few signals of the non-zero entries in $\boldsymbol{\beta}$ are weaker. Here, we fix $q=5$, and the true value of $\boldsymbol{\beta}$ is $\boldsymbol{\beta}_0= \{ 1,-0.5,0,\ldots,0,-1,0.4,0.75 \}^\top$ in this case. The true values of $\boldsymbol{\alpha}$ and the non-linear functions are the same as in Scenario 1. 
\begin{table}
\centering
\caption{Variable selection and estimation results over 200 replications for Scenario 2. Standard deviations of the MMSE are in parentheses.} \label{WS}
\begin{tabular}{l|llllll}
\hline
Method & MMSE & TP & FP & MS & MC & TM \\
\hline
\multicolumn{7}{c}{$n=600,p=300$} \\
\hline
BAR(AIC) & 0.19(0.13) & 4.63 & 1.28 & 5.91 & 1.65 & 21$\%$ \\
BAR(BIC) & 0.50(0.18) & 3.14 & 0 & 3.14 & 1.86 & 0$\%$ \\
LASSO & 0.67(0.15) & 4.66 & 6.51 & 11.17 & 6.85 & 3$\%$  \\
ALASSO & 0.33(0.18) & 4.59 & 6.52 & 11.11 & 6.93 & 8$\%$  \\
Oracle & 0.07(0.05) & 5 & 0 & 5 & 0 & 100$\%$ \\
\hline
\multicolumn{7}{c}{$n=800,p=300$} \\
\hline
BAR(AIC) & 0.13(0.09) & 4.82 & 1.32 & 6.14 & 1.50 & 24$\%$ \\
BAR(BIC) & 0.41(0.14) & 3.46 & 0 & 3.46 & 1.54 & 5$\%$ \\
LASSO & 0.52(0.14) & 4.85 & 8.06 & 12.91 & 8.21 & 2$\%$   \\
ALASSO & 0.24(0.11) & 4.81 & 5.93 & 10.74 & 6.12 & 14$\%$  \\
Oracle & 0.05(0.05) & 5 & 0 & 5 & 0 & 100$\%$ \\
\hline
\multicolumn{7}{c}{$n=600,p=450$} \\
\hline
BAR(AIC) & 0.23(0.14) & 4.65 & 1.89 & 6.54 & 2.24 & 8$\%$  \\
BAR(BIC) & 0.50(0.16) & 3.22 & 0 & 3.22 & 1.78 & 1$\%$ \\
LASSO & 0.68(0.18) & 4.65 & 8.66 & 13.31 & 9.01 & 2$\%$  \\
ALASSO & 0.34(0.23) & 4.54 & 10.20 & 14.74 & 10.66 & 2$\%$  \\
Oracle & 0.07(0.07) & 5 & 0 & 5 & 0 & 100$\%$ \\
\hline
\multicolumn{7}{c}{$n=800,p=450$} \\
\hline
BAR(AIC) & 0.15(0.09) & 4.87 & 1.90 & 6.77 & 2.03 & 10$\%$  \\
BAR(BIC) & 0.41(0.13) & 3.48 & 0 & 3.48 & 1.52 & 5$\%$ \\
LASSO& 0.52(0.14) & 4.87 & 9.15 & 14.02 & 9.28 & 1$\%$  \\
ALASSO & 0.23(0.12) & 4.79 & 9.11 & 13.90 & 9.32 & 6$\%$  \\ 
Oracle & 0.05(0.04) & 5 & 0 & 5 & 0 & 100$\%$ \\
\hline
\end{tabular}
\end{table}
One is able to observe the GPLM-BAR method with the AIC penalty outperforms the LASSO and Adaptive LASSO methods from the results in Table \ref{WS}. However, in comparison to the results in Scenario 1, the selection and estimation accuracy become worse, because the weaker signals in $\boldsymbol{\beta}$ have a greater tendency to be shrunk to zero. The estimation of the non-linear covariate effects using the GPLM-BAR method are good (Figures \ref{ESTS2p300n600},\ref{ESTS2p300n800},\ref{ESTS2p450n600} and \ref{ESTS2p450n800}), and the estimation results of $\boldsymbol{\alpha}$ are satisfactory (Table \ref{alphaS2}).

\subsection*{Scenario 3: CATHGEN-based simulations}

By using the results in the real data analysis, we also perform a simulation study to investigate the performance of our method under the correlation structure of the CATHGEN data. The results are presented in the Appendix.

\subsection*{Scenario 4: Poisson partly linear model}

We also perform a simulation study for the Poisson partly linear model. The simulated results are presented in the Appendix. 

\section{Real data analysis} \label{rda}
Coronary artery disease (CAD) is a major disease that inflicts death, and is one of the biggest causes of death globally \citep{abubakar2015global}. Environmental factors that contribute to CAD are typically age, smoking status, obesity and lifestyle choices. However, genetic factors play a role in death due to CAD, especially in younger patients \citep{marenberg1994genetic}. 

We apply our proposed method on the CATHGEN data, which was downloaded from dbGaP, with accession number phs000704.v1.p1. The study collected peripheral blood samples from consenting patients who were undergoing cardiac catheterization at Duke University Medical Center from 2001 to 2011. A total of 1327 patients were recruited and followed-up between 2004 until 2014. The binary response variable is the affection status, where the stratification criteria is defined in \cite{shah2010reclass}. The high-dimensional design matrix contains 13991 columns of SNPs belonging to 331 genes that have been associated with CAD using Ingenuity Pathway Analysis \citep{kramer2014causal}. In addition to the SNPs, there are ten clinical and demographical variables in the data. These variables include age (Mean = 57.0, SD = 11.6), BMI (Mean = 30.8, SD = 7.8), smoking status (671 cases out of 1327), race (897 Caucasian-Americans, 274 African-Americans and 156 Asian-Americans), hypertension status (900 cases out of 1327), diabetes status (379 cases out of 1327), hypercholesterolemia status (745 cases out of 1327), sex (684 males and 643 females), number of diseased vessels and history of myocardial infarction (HXMI) (277 cases out of 1327). All clinical and demographical variables of each subject were measured when they are included to the study. We exclude the number of diseased vessels from further analysis because of conversion issues when fitting the univariate logistic regression model.

\begin{figure}
\centering
\centerline{\includegraphics[height = 8cm, width = 10cm]{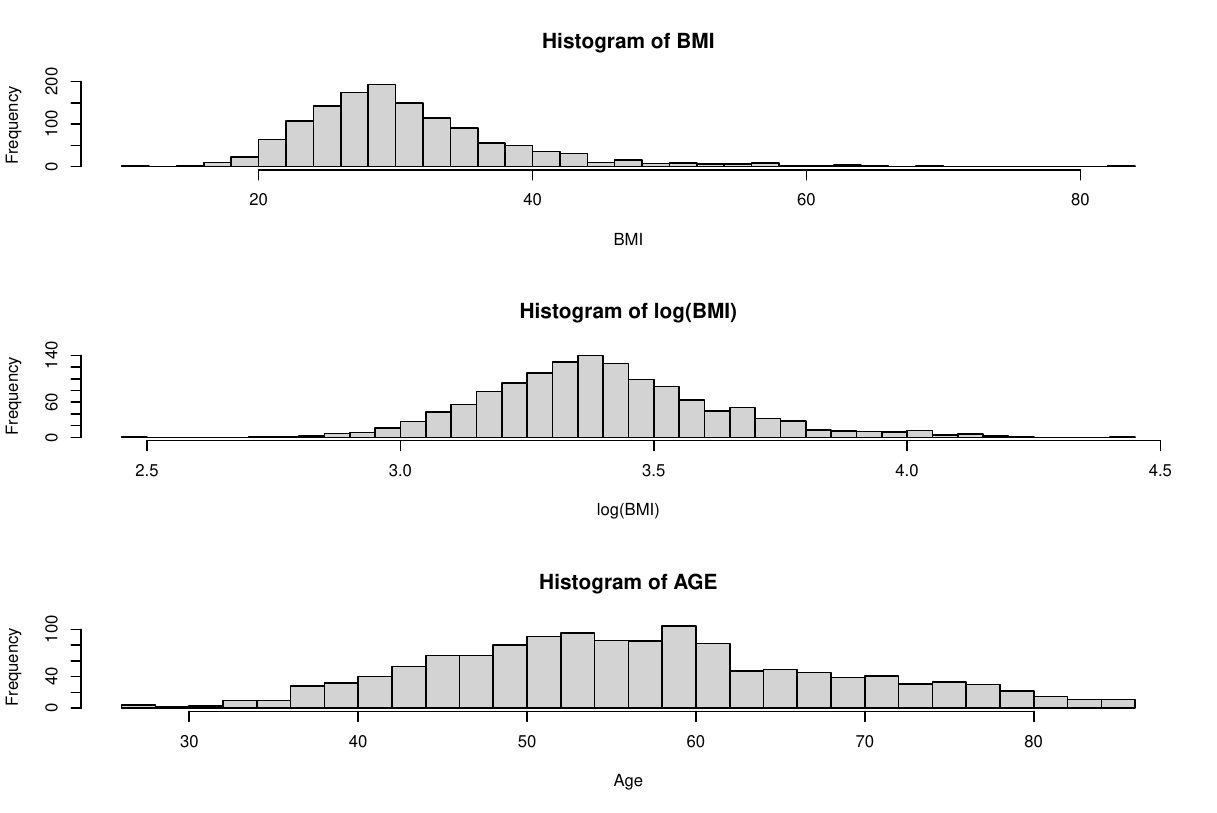}}
\caption{The histograms of the BMI on the original scale (top panel), BMI on the log-transformed scale (middle panel), and the histogram of age on the original scale (bottom panel). } \label{HIST}
\end{figure}

\begin{table}
\centering
\caption{Univariate polynomial logistic regression models of age and $\log$(BMI) fitted separately, with their respective higher order terms.} \label{Uni}
\begin{tabular}{l|llll}
\hline
Variable & Estimate & Std. Error & z-value & p-value \\
\hline
Intercept & -3.575 & 1.168 & -3.061 & 2E-03\\
Age & 0.107 & 4E-02 & 2.645 & 8E-03 \\
$\text{Age}^2$ & -8E-04 & 4E-04 & -2.220 & 2.6E-02 \\
\hline
Intercept & -38.952 & 8.684 &-4.486 & 7.3E-06 \\
$\log(\text{BMI})$ & 22.561 & 5.035 & 4.480 & 7.45E-06 \\
$\log(\text{BMI})^2$ & -3.255 &0.729 & -4.468 & 7.88E-06 \\
\hline
\end{tabular}
\end{table}
In Figure \ref{HIST}, the distribution of age is symmetrical on the original scale. However, the distribution of the BMI on the original scale is right skewed. The natural logarithm transformation of it fixes the skewness. Thus, we decide to use the BMI on the log scale for further analysis. From Table \ref{Uni}, when fitting age and it's second order polynomial term in the logistic regression model, both terms are found to be statistically significant. Likewise, when fitting the log-transformed BMI and its second order polynomial term in the logistic regression model, both terms are also significant. The results in Table \ref{Uni} indicate that age and log-transformed BMI have a non-linear effect on the odds ratio of developing CAD. However, the functional form of the effect is unknown, and this motivates us to consider a logistic partly linear regression model. 

Before we apply our proposed method, it is clear that the dimension of the design matrix needs to be reduced. To reduce the dimension, we first remove SNPs with a minor allele frequency (MAF) of less than 0.1. We then further reduce the number of SNPs through pre-screening the candidate SNPs, by performing univariate logistic regression, only selecting SNPs with a $p$-value less than 0.1. In total, 1242 SNPs with a $p$-value less than 0.1 are retained for further analysis.

\begin{figure}
\centering
\includegraphics[height=6.75cm, width=9cm]{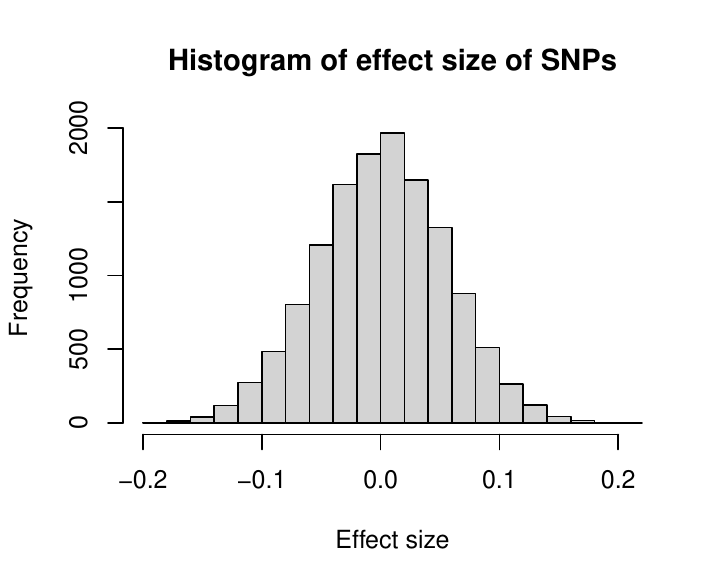}
\caption{Histogram of the estimated effect sizes of SNPs using univariate logistic regression.} \label{fig4}
\end{figure}

To choose the tuning parameters in GPLM-BAR, we decide to use the AIC penalty for $\lambda_n$, because the individual estimated effect sizes of the SNPs are small as shown in Figure \ref{fig4}. The value of $\xi_n$ and number of basis functions are kept the same as the simulation study. The tuning parameters values for the LASSO and Adaptive LASSO methods are chosen by 5-fold cross validation. In Table \ref{table4}, the estimated effects of the categorical clinical variables obtained from the GPLM-BAR method have a larger magnitude. The results in Table \ref{table4} indicate a positive risk association for hypertension, diabetes, hypercholesterolomia, smoking and HXMI, where HXMI is the strongest clinical indicator on the risk of developing CAD. We use the bootstrap method with 100 random bootstrap samples to obtain the estimated standard error in parentheses in Table \ref{table4}. The GPLM-BAR method identified the fewest number of SNPs that contribute to CAD. Specifically, the GPLM-BAR identified 19 different SNPs that are associated to 17 unique genes, the LASSO and Adaptive LASSO methods identified 199 SNPs and 228 SNPs, respectively. 

\begin{table}
\centering
\caption{Estimation results of the categorical clinical variables for the CATHGEN data.} \label{table4}
\begin{tabular}{l| lll }
\hline
Variable & GPLM-BAR & LASSO & ALASSO \\
\hline
Hypertension & $0.368_{(0.315)}$ & $0.324_{(0.157)}$ & $0.259_{(0.093)}$\\
Diabetes & $1.180_{(0.269)}$ & $1.097_{(0.242)}$ & $1.105_{(0.261)}$\\
Hypercholesterolomia & $1.403_{(0.299)}$ & $1.459_{(0.275)}$ & $1.399_{(0.234)}$ \\
Sex & $0.361_{(0.224)}$ & $0.343_{(0.193)}$ & $0.313_{(0.219)}$\\
Smoking & $0.860_{(0.335)}$ & $0.787_{(0.301)}$ & $0.797_{(0.221)}$ \\
HXMI & $37.388_{(1.786)}$ & $10.620_{(0.745)}$ & $10.477_{(0.561)}$ \\
Race (African) & $-0.030_{(0.489)}$ & $-0.163_{(0.341)}$ & $0.879_{(0.584)}$\\
Race (Caucasian) & $0.385_{(0.458)}$ & $0.544_{(0.528)}$ & $0.084_{(0.394)}$\\
\hline 
\end{tabular}
\caption*{Male and Asian-American are the reference categories for the variables Sex and Race, respectively. The remaining variables have the non-cases as the reference categories.}
\end{table}

From the genes identified using GPLM-BAR, \textit{RBFOX1} is found to be associated with blood pressure and heart failure through transcriptome profiling \citep{gao2016rbfox1}. \textit{CDH13} has been shown to be associated with blood cholesterol and CAD through a genome-wide association study undertaken in the British population \citep{wellcome2007genome}.  \textit{F10} is associated with the lowering levels of coagulation factor X, which is protective against ischemic heart disease \citep{paraboschi2020rare}. \textit{GABRG3} has been shown to be associated with density of dodecanedioic acid, which plays a role in regulating blood sugar level \citep{wang2021genome}. \textit{ABCA1} has been shown to be associated to altered lipoprotein levels which results in a increased risk for CAD \citep{clee2001common}. \textit{IL1B} belongs to the wider family of \textit{IL1} genes which is associated to coronary heart disease \citep{francis1999interleukin,vohnout2003interleukin,tsimikas2014pro}. Certain subtypes of the \textit{APOE} gene are identified to lipid levels and coronary risk \citep{bennet2007association}. We report the complete set of selected SNPs and genes in Table \ref{SNPgene}. 

\begin{figure}
\centering
\includegraphics[width =11.8cm, height =7.5cm]{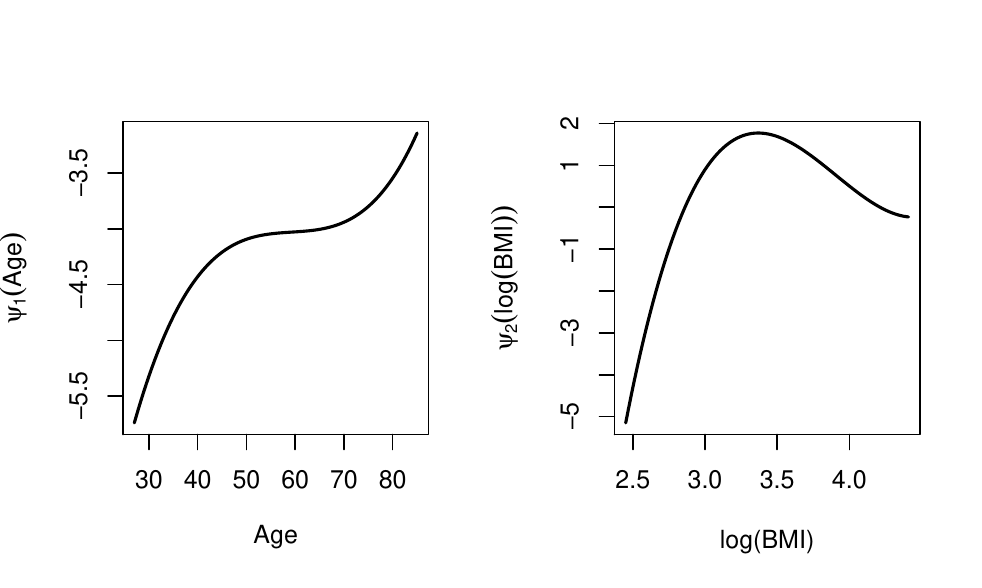}
\caption{Estimated covariate effects of age (left panel) and log(BMI) (right panel).} \label{phiagebmi}
\end{figure}

In addition to the results in Table \ref{table4}, one can observe our method using Bernstein polynomials approximation has showed that the effects of age and BMI are non-linear, as seen in Figure \ref{phiagebmi}. The plot on the left in Figure \ref{phiagebmi} shows the risk of developing CAD increases non-linearly with age, and the plot on the right shows the risk of developing CAD increases with BMI on the natural logarithm scale until 3.5. After this cutoff point, it then decreases. The unusual trend seen for BMI can be partially explained by the lack of data when $\text{BMI} > 35$ or $\log(\text{BMI}) > 3.5$, as BMI of the majority of patients recruited to this study falls between 15 and 35 on the raw scale.

\section{Discussion and Conclusion} \label{sec:diss}
In this article, we have proposed a new approach for simultaneous variable selection and estimation under the context of GPLM, with a focus on the logistic partly linear regression model. Our proposed approach was motivated by the CATHGEN study, where the data contains both high-dimensional genetic covariates and low-dimensional clinical and demographical covariates. We considered GPLM as it grants us the flexibility to model possible non-linear covariate effects. We employed the Bernstein polynomials to approximate the non-parametric component of the model, where it has several advantages over other approximation methods like splines and piecewise functions. First, unlike the piecewise functions the Bernstein polynomials are differentiable and continuous everywhere. This is desirable as the first and second derivatives are calculated in each iteration of our algorithm. Second, the Bernstein polynomials possesses computational scalability and optimal shape-preserving property for all approximating polynomials \citep{carnicer1993shape}. Third, the Bernstein polynomials do not require specification of the number of interior knots and their locations, unlike B-splines. From the results of our comprehensive simulation studies, we observe that our proposed method outperforms common variable selection methods under a few practical scenarios. Our method incorporating the BAR penalty produced the lowest total misclassification rate and the highest frequency of the true model selected, which is consistent with other empirical studies conducted by authors who also employed the BAR penalty\citep{dai2018broken,zhao2019simultaneous,sun2022broken}. Our method was also able to accurately estimate the true non-linear functions. As an application, we applied our proposed method to the CATHGEN data, where certain SNPs and genes were found to have a relevant contribution to CAD, which is consistent with other variable selection methods applied to this data \citep{li2022bayesian, dai2023bayesian}. 
 
There are several directions one can take from our research. In the simulation study, we only examined scenarios where the number of high-dimensional covariates $p$ are diverging with a rate less than the sample size $n$. Suppose $p$ diverges with a rate greater than $n$, it would be of interest to investigate the performance of our method under this scenario. The CATHGEN data also contains right-censored survival information. Under the context of survival models, it would be of interest to investigate which relevant genetic markers affect the survival probability, and the possible homogeneity or heterogeneity of the two sets of genetic markers selected based on the two different response outcomes and their biological interpretations. Additionally, choosing the optimal tuning parameter poses a significant challenge to researchers. The mixture of weak signals with strong signals poses a noteworthy problem to researches, and requires more thorough investigation.

\section*{Acknowledgement}
The authors would like to acknowledge New Frontiers in Research Fund to Quan Long (NFRFE-2018-00748) administered by the Canada Research Coordinating Committee, the Canada Foundation for Innovation JELF grant (36605) awarded to Quan Long, and the Discovery Grant to Xuewen Lu (RG/PIN06466-2018) administered by the National Science and Engineering Research Council of Canada. 

\section*{Declaration of Conflicting Interests} 

The author(s) declared no potential conflicts of interest with respect to the research, authorship, and/or publication of this article.

\bibliographystyle{plainnat}
\bibliography{arXiVOct312023}
\newpage

\section*{Appendix}
\subsection*{Cyclic coordinate descent algorithm}
The cyclic coordinate descent algorithm first sets all parameters to some chosen initial value. It solves a one-dimensional optimization problem by estimating the first parameter that minimizes the objective function, while holding the other parameters constant. It then estimates subsequent parameters by solving a one-dimensional problem and holding the other parameters constant. When all parameters are estimated, the iteration is complete and returns to the first parameter for the algorithm to be repeated. Multiple iterations are done over the whole set of parameters until the pre-specified convergence criteria is met. Computing the Hessian matrix or gradient vector, inverting the Hessian matrix, and transposing the gradient vector are not required in the CLG algorithm, as only one-dimensional updates are used. As a result, the CLG algorithm easily scales to high-dimensional data \citep{wu2008coordinate, simon2011regularization, gorst2012coordinate}. It has then been implemented under GLM for massive sample size data \citep{suchard2013massive, li2021scalable} and high-dimensional massive sample size Cox PH model \citep{mittal2014high}. \citet{suchard2013massive} and \citet{mittal2014high} developed an R package \verb|Cyclops| that incorporates ridge and LASSO regularization using Gaussian and Laplace priors. More specifically, when $s=0$, the following prior is needed to obtain the initial estimates $\widehat{\boldsymbol{\beta}}^{(0)}: f(\beta_j | \xi_n) \sim N(0, 1/ \xi_n)$. For each $s \geq 1$, the reweighted prior $f(\widehat{\beta}^{(s)}_j | \lambda_n) \sim N(0, \widehat{\beta}^{(s-1)2}_j/\lambda_n)$ is used to obtain $\widehat{\boldsymbol{\beta}}^{(s)}$.

The CLG algorithm finds $\{\alpha^{(\text{new})}_j,\beta^{(\text{new})}_j,\gamma^{(\text{new})}_{jk} \}$, where $\{\alpha^{(\text{new})}_j,\beta^{(\text{new})}_j \}$ are the updated estimates of the $j^{th}$ entry of $\boldsymbol{\alpha}$ and $\boldsymbol{\beta}$, and $\gamma^{(\text{new})}_{jk}$ is the updated estimated of the $(j,k)^{th}$ entry of $\boldsymbol{\gamma}$, while keeping the other values of $\alpha_j$'s,$\beta_j$'s and $\gamma_{jk}$'s constant. Suppose we have a tuning parameter $1/\phi_j$. Therefore, when $s \geq 0$, finding $\beta^{(\text{new})}_j$ for $j=1,\ldots,p$ is equivalent to finding $u$ that minimizes the penalized negative log-likelihood
\begin{equation}\label{CLG}
\begin{split}
f(u) = & - \sum^n_{i=1} \bigg\{ y_i \Big(u x_{ij} + \sum^p_{k=1, j \neq k} \beta_k x_{ik} + \sum^{q_w}_{k=1} \alpha_k w_{ik} + \sum^{q_z}_{k=1} \sum^{m_k}_{l=0} \gamma_{kl} B_{kl}(z_{ik},m_k,c_k,u_k)\Big) \\
  &- \log \Big(1+\exp\big\{ u x_{ij} + \sum^p_{k=1, j \neq k} \beta_k x_{ik} + \sum^{q_w}_{k=1} \alpha_k w_{ik} + \sum^{q_z}_{k=1} \sum^{m_k}_{l=0} \gamma_{kl} B_{kl}(z_{ik},m_k,c_k,u_k) \big\} \Big) \bigg\} \\
  & + \frac{u^2}{2\phi_j}.
\end{split}
\end{equation}
At the current $\beta_j$, one could use the Taylor expansion to approximate $f(u)$ by
$$
f(u) \approx f(\beta_j) + f'(\beta_j)(u-\beta_j) + \frac{1}{2}f''(\beta_j)(u-\beta_j)^2.
$$ 
In \eqref{CLG}, different penalty terms can used. For example, $\phi_j = 1/\xi_n$ and $\phi_j = (\widehat{\beta}^{(s-1)}_j)^2/ \lambda_n$ in the GPLM-BAR algorithm. Similarly, finding $\alpha^{(\text{new})}_j$ for $j=1,\ldots,q_w$ is equivalent to minimizing
\begin{equation*}
\begin{split}
g(u) = & - \sum^n_{i=1} \bigg\{ y_i \Big(\sum^p_{k=1} \beta_k x_{ik} + u w_{ij} + \sum^{q_w}_{k=1, j \neq k} \alpha_k w_{ik} + \sum^{q_z}_{k=1} \sum^{m_k}_{l=0} \gamma_{kl} B_{kl}(z_{ik},m_k,c_k,u_k) \Big) \\
& - \log\Big(1+\exp\big\{ \sum^p_{k=1} \beta_k x_{ik} + u w_{ij} + \sum^{q_w}_{k=1, j \neq k} \alpha_k w_{ik} + \sum^{q_z}_{k=1} \sum^{m_k}_{l=0} \gamma_{kl} B_{kl}(z_{ik},m_k,c_k,u_k) \big\}\Big) \bigg\},
\end{split}
\end{equation*}
and finding $\gamma^{(\text{new})}_{jk}$ for $j=1,\ldots,q_z$ and $k=1,\ldots,m_j$ is equivalent to minimizing
\begin{equation*}
\begin{split}
h(u) & = - \sum^n_{i=1} \Bigg\{ y_i \Big(\sum^p_{k=1} \beta_k x_{ik} + \sum^{q_w}_{k=1} \alpha_k w_{ik} + uB_{jk}(z_{ij}, m_j, c_j, u_j) + \sum^{q_z}_{s=1, j \neq s} \sum^{m_s}_{t=0, k \neq t} \gamma_{st} B_{st}(z_{is},m_s,c_s,u_s)   \Big)  \\
& - \log\bigg(1+\exp\Big\{ \sum^p_{k=1} \beta_k x_{ik} + \sum^{q_w}_{k=1} \alpha_k w_{ik} + uB_{jk}(z_{ij}, m_j, c_j, u_j) + \sum^{q_z}_{s=1, j \neq s} \sum^{m_s}_{t=0, k \neq t} \gamma_{st} B_{st}(z_{is},m_s,c_s,u_s) \Big\} \bigg) \Bigg\}.
\end{split}
\end{equation*}
Suppose the Taylor series is also used to approximate $g(u)$ and $h(u)$, then values of $\{\alpha^{(\text{new})}_j,\beta^{(\text{new})}_j,\gamma^{(\text{new})}_{jk} \}$ can be computed as
$$
\beta^{(\text{new})}_j = \beta_j - \frac{f'(\beta_j)}{f''(\beta_j)}, \;\; \alpha^{(\text{new})}_j = \alpha_j - \frac{g'(\alpha_j)}{g''(\alpha_j)}, \;\; \gamma^{(\text{new})}_{jk} = \gamma_{jk} - \frac{h'(\gamma_{jk})}{h''(\gamma_{jk})}.
$$
The first-order and second-order derivations of $f(u)$, $g(u)$ and $h(u)$ are computed as follows. When $s=0$, $\phi_j = 1/\xi_n$, then the first-order and second-order derivatives are
\begin{equation*}
\begin{split}
f'(u) & = \left. \frac{\partial}{\partial u} f(u) \right\vert_{u=\beta_j}  \\ 
& = - \sum^n_{i=1} \left\{ x_{ij}y_i - \frac{x_{ij} \exp\{u x_{ij} + \sum^p_{k=1, j \neq k} \beta_k x_{ik} + \sum^{q_w}_{k=1} \alpha_k w_{ik} + \sum^{q_z}_{k=1} \sum^{m_k}_{l=0} \gamma_{kl} B_{kl}(z_{ik},m_k,c_k,u_k) \}}{1 + \exp\{u x_{ij} + \sum^p_{k=1, j \neq k} \beta_k x_{ik} + \sum^{q_w}_{k=1} \alpha_k w_{ik} + \sum^{q_z}_{k=1} \sum^{m_k}_{l=0} \gamma_{kl} B_{kl}(z_{ik},m_k,c_k,u_k) \}}\right\} \\
& + 2\xi_n u,
\end{split}
\end{equation*}
and
\begin{equation*}
\begin{split}
f''(u) & = \left. \frac{\partial^2}{\partial u^2} f(u) \right\vert_{u=\beta_j} \\
& = - \sum^n_{i=1} \left\{ -\frac{x^2_{ij} \exp\{u x_{ij} + \sum^p_{k=1, j \neq k} \beta_k x_{ik} + \sum^{q_w}_{k=1} \alpha_k w_{ik} + \sum^{q_z}_{k=1} \sum^{m_k}_{l=0} \gamma_{kl} B_{kl}(z_{ik},m_k,c_k,u_k)\}}{(1 + \exp\{u x_{ij} + \sum^p_{k=1, j \neq k} \beta_k x_{ik} + \sum^{q_w}_{k=1} \alpha_k w_{ik} + \sum^{q_z}_{k=1} \sum^{m_k}_{l=0} \gamma_{kl} B_{kl}(z_{ik},m_k,c_k,u_k)\})^2}\right\} \\
& + 2\xi_n,
\end{split}
\end{equation*}
respectively. When $s \geq 1$, $\phi_j = (\widehat{\beta}^{(s-1)}_j)^2/\lambda_n$, then
\begin{equation*}
\begin{split}
f'(u) & = \left. \frac{\partial}{\partial u} f(u) \right\vert_{u=\beta_j} \\ &= - \sum^n_{i=1} \bigg\{x_{ij}y_i - \frac{x_{ij} \exp\{u x_{ij} + \sum^p_{k=1, j \neq k} \beta_k x_{ik} + \sum^{q_w}_{k=1} \alpha_k w_{ik} + \sum^{q_z}_{k=1} \sum^{m_k}_{l=0} \gamma_{kl} B_{kl}(z_{ik},m_k,c_k,u_k) \}}{1 + \exp\{u x_{ij} + \sum^p_{k=1, j \neq k} \beta_k x_{ik} + \sum^{q_w}_{k=1} \alpha_k w_{ik} +\sum^{q_z}_{k=1} \sum^{m_k}_{l=0} \gamma_{kl} B_{kl}(z_{ik},m_k,c_k,u_k) \} }\bigg\} \\
& + \frac{2\lambda_n u}{(\widehat{\beta}^{(s-1)}_j)^2},
\end{split}
\end{equation*}
and
\begin{equation*}
\begin{split}
f''(u) & = \left. \frac{\partial^2}{\partial u^2} f(u) \right\vert_{u=\beta_j} \\  &= - \sum^n_{i=1} \left\{\
- \frac{x^2_{ij} \exp\{u x_{ij} + \sum^p_{k=1, j \neq k} \beta_k x_{ik} + \sum^{q_w}_{k=1} \alpha_k w_{ik} + \sum^{q_z}_{k=1} \sum^{m_k}_{l=0} \gamma_{kl} B_{kl}(z_{ik},m_k,c_k,u_k) \}}{(1 + \exp\{u x_{ij} + \sum^p_{k=1, j \neq k} \beta_k x_{ik} + \sum^{q_w}_{k=1} \alpha_k w_{ik} + \sum^{q_z}_{k=1} \sum^{m_k}_{l=0} \gamma_{kl} B_{kl}(z_{ik},m_k,c_k,u_k) \})^2} \right\} \\
& + \frac{2\lambda_n}{(\widehat{\beta}^{(s-1)}_j)^2}.
\end{split}
\end{equation*}
Similarly, the first and second derivatives of $g(u)$ are
\begin{equation*}
\begin{split}
g'(u) & = \left. \frac{\partial}{\partial u} g(u) \right\vert_{u=\alpha_j} \\ & = - \sum^n_{i=1} \bigg\{w_{ij} y_i \\
&- \frac{w_{ij} \exp\{ \sum^p_{k=1} \beta_k x_{ik} + + u w_{ij} + \sum^{q_w}_{k=1, j \neq k} \alpha_k w_{ik} + \sum^{q_z}_{k=1} \sum^{m_k}_{l=0} \gamma_{kl} B_{kl}(z_{ik},m_k,c_k,u_k) \} }{1+\exp\{ \sum^p_{k=1} \beta_k x_{ik} + u w_{ij} + \sum^{q_w}_{k=1, j \neq k} \alpha_k w_{ik} +  \sum^{q_z}_{k=1} \sum^{m_k}_{l=0} \gamma_{kl} B_{kl}(z_{ik},m_k,c_k,u_k) \}} \bigg\},
\end{split}
\end{equation*}
and
\begin{equation*}
\begin{split}
g''(u) & = \left. \frac{\partial^2}{\partial u^2} g(u) \right\vert_{u=\alpha_j} \\ & = - \sum^n_{i=1} \bigg\{- \frac{w^2_{ij} \exp\{ \sum^p_{k=1} \beta_k x_{ik} + u w_{ij} +\sum^{q_w}_{k=1, j \neq k} \alpha_k w_{ik} + \sum^{q_z}_{k=1} \sum^{m_k}_{l=0} \gamma_{kl} B_{kl}(z_{ik},m_k,c_k,u_k) \} }{(1+\exp\{ \sum^p_{k=1} \beta_k x_{ik} + u w_{ij} +\sum^{q_w}_{k=1, j \neq k} \alpha_k w_{ik} +  \sum^{q_z}_{k=1} \sum^{m_k}_{l=0} \gamma_{kl} B_{kl}(z_{ik},m_k,c_k,u_k) \})^2} \bigg\}.
\end{split}
\end{equation*}
Finally, the first and second derivatives of $h(u)$ are
\begin{equation*}
\begin{split}
h'(u) & = \left. \frac{\partial}{\partial u} h(u) \right\vert_{u=\gamma_{jk}} \\
& = - \sum^n_{i=1} \bigg\{ \Ddot{B}_{jk} y_i  - \frac{\Ddot{B}_{jk} \exp\{ \sum^p_{k=1} \beta_k x_{ik} + \sum^{q_w}_{k=1} \alpha_k w_{ik} + u \Ddot{B}_{jk} + \sum^{q_z}_{s=1, j \neq s} \sum^{m_s}_{t=0, k \neq t} \gamma_{st} \Ddot{B}_{st}\}}{1+\exp\{ \sum^p_{k=1} \beta_k x_{ik} + \sum^{q_w}_{k=1} \alpha_k w_{ik} + u \Ddot{B}_{jk} + \sum^{q_z}_{s=1, j \neq s} \sum^{m_s}_{t=0, k \neq t} \gamma_{st} \Ddot{B}_{st}\}}\bigg \} ,
\end{split}
\end{equation*}
and
\begin{equation*}
\begin{split}
h''(u) &= \left. \frac{\partial^2}{\partial u^2} h(u) \right\vert_{u=\gamma_{jk}} \\ = & - \sum^n_{i=1} \bigg\{ - \frac{\Ddot{B}^2_{jk} \exp\{ \sum^p_{k=1} \beta_k x_{ik} + \sum^{q_w}_{k=1} \alpha_k w_{ik} + u \Ddot{B}_{jk} + \sum^{q_z}_{s=1, j \neq s} \sum^{m_s}_{t=0, k \neq t} \gamma_{st} \Ddot{B}_{st} \}}{(1+\exp\{ \sum^p_{k=1} \beta_k x_{ik} + \sum^{q_w}_{k=1} \alpha_k w_{ik} + u \Ddot{B}_{jk} + \sum^{q_z}_{s=1, j \neq s} \sum^{m_s}_{t=0, k \neq t} \gamma_{st} \Ddot{B}_{st} \})^2}\bigg\} ,
\end{split}
\end{equation*}
where $\Ddot{B}_{jk}= B_{jk}(z_{ik},m_k,c_k,u_k)$ and $\Ddot{B}_{st} = B_{st}(z_{is}, m_s, c_s, u_s)$.

\section*{Additional results of the simulation studies}

\subsection*{Scenario 3: CATHGEN-based simulations}
We perform a simulation study based on the CATHGEN data, where the motivation is to examine the performance of our proposed method, in comparison with existing methods, under the correlation structure of our data. Before conducting the simulation study, we first have to screen the data to reduce dimension. The final screened set is 1313 observations and 1242 SNPs. Then, we use the GPLM-BAR to determine which SNPs are relevant to the binary response, where 19 SNPs are selected. In the simulation set-up, we set the effect size of the identified SNPs as four times the true estimated effect size. We do this because the original effects are too small for the model to distinguish them. We do not change the effect size of the categorical and continuous variables. We also do not change the coefficients of the Bernstein polynomials.

\begin{figure}
    \centering
    \includegraphics[height=8cm, width=10.5cm]{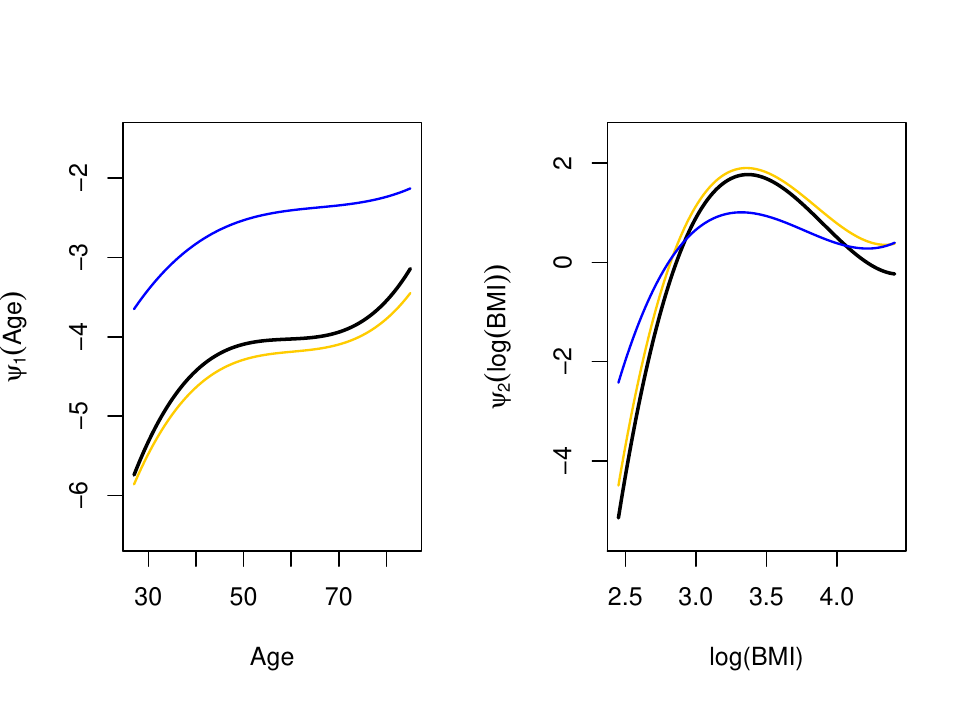}
    \caption{Estimation of the non-linear effects of age and log(BMI) based on the simulated CATHGEN data. The black curve represents the true effect, yellow curve represents the GPLM-BAR with AIC penalty, and the blue curve represents the GPLM-BAR with BIC penalty.} \label{S3plot}
\end{figure}

\begin{table}[H]
\centering
\caption{Variable selection and estimation results over 200 replications for Scenario 3. Standard deviation of the MMSE are in parentheses.} \label{S3}
\begin{tabular}{l |llllll}
\hline
Method & MMSE & TP & FP & MS & MC & TM \\
\hline
BAR(AIC) & 0.57(0.25) & 18.84 & 2.22 & 21.06 & 2.38 & 9$\%$ \\
BAR(BIC) & 6.75(0.98)& 12.41 & 0.07 & 12.48 & 6.66 & 0$\%$\\
LASSO & 5.24(0.65) & 18.97 & 52.73 & 71.70 & 52.76 & 0$\%$\\
ALASSO & 2.82(0.63) & 18.89 & 27.48 & 46.37 & 27.59 &  0$\%$\\
Oracle & 0.30(0.26) & 19 & 0 & 19 & 0 & 100$\%$\\
\hline
\end{tabular}
\end{table}
From Table \ref{S3}, the GPLM-BAR method with AIC penalty has the lowest average number of FP and MC. Using the GPLM-BAR with BIC penalty, the average number of TP decreases due to the large size of the tuning parameter $\lambda_n$. The Lasso and Adaptive Lasso methods produce very large false positives. In Table \ref{S3Est}, the GPLM-BAR with AIC penalty produces the least biased estimates of the relevant SNPs. In Figure \ref{S3plot}, we observe that the estimation of the non-linear effects of age and log(BMI) is also better when the AIC penalty, as compared to the BIC penalty, is used. 
\begin{table}[H]
    \centering
    \caption{Averaged estimates of the relevant SNPs identified in the real data analysis, performed over 200 replication for Scenario 3, along with the clinical and demographical variables. Standard deviations are in parentheses.} \label{S3Est}
    \centerline{
    \begin{tabular}{c|c|ccccc}
    \hline
     & True & BAR-AIC & BAR-BIC & LASSO & ALASSO & Oracle\\
    \hline     
    rs1407961 & -1.024 & -0.922(0.136) & -0.444(0.134) & -0.432(0.075) & -0.630(0.099) & -1.088(0.144)\\
    rs8037353 & 1.017 & 0.898(0.121) & 0.277(0.192) & 0.376(0.072) & 0.567(0.098) & 1.099(0.157)\\
    rs7107322 & 1.015 & 0.909(0.129) & 0.300(0.176) & 0.387(0.078) & 0.560(0.105) & 1.097(0.139)\\
    rs2387952 & -1.336 & -1.228(0.129) & -0.709(0.077) & -0.621(0.081) & -0.843(0.102) & -1.410(0.153)\\
    rs6680365 & -0.918 & -0.830(0.118) & -0.370(0.131) & -0.401(0.069) & -0.555(0.096) & 0.980(0.131)\\
    rs3136558 & -0.906 & -0.806(0.116) & -0.363(0.154) & -0.372(0.067) & -0.530(0.091) & 0.961(0.135)\\
    rs4131888 & 0.512 & 0.419(0.137) & 0.008(0.049) & 0.180(0.070) & 0.253(0.106) & 0.543(0.118)\\
    rs3845439 & 1.106 & 0.936(0.276) & 0.411(0.154) & 0.366(0.123) & 0.533(0.168) & 1.182(0.136)\\
    rs769449 & 0.773 & 0.701(0.104) & 0.200(0.174) & 0.338(0.063) & 0.479(0.083) & 0.833(0.125)\\
    rs9549675 & -1.092 & -0.968(0.127) & -0.461(0.109) & -0.438(0.075) & -0.636(0.100) & -1.169(0.157)\\
    rs2805543 & -1.203 & -1.091(0.123) & -0.538(0.093) & -0.531(0.085) & -0.731(0.106) & -1.275(0.144)\\
    rs821292 & -0.781 & -0.666(0.136) & -0.033(0.096) & -0.247(0.076) & -0.362(0.117) & -0.840(0.119)\\
    rs12612481 & 0.770 & 0.698(0.116) & 0.266(0.161) & 0.329(0.075) & 0.448(0.104) & 0.824(0.108)\\
    rs7188981 & 0.905 & 0.823(0.120) & 0.402(0.136) & 0.398(0.074) & 0.560(0.101) & 0.975(0.128)\\
    rs9932172 & 0.997 & 0.903(0.120) & 0.483(0.083) & 0.435(0.075) & 0.591(0.104) & 1.044(0.138)\\
    rs9282537 & 0.763 & 0.658(0.129) & 0.045(0.117) & 0.256(0.074) & 0.358(0.119) & 0.826(0.123)\\
    rs244072 & -0.608 & -0.509(0.113) & 0 & -0.162(0.066) & -0.234(0.103) & -0.660(0.136)\\
    rs11859718 & 0.760 & 0.675(0.138) & 0.279(0.176) & 0.316(0.071) & 0.440(0.010) & 0.811(0.136)\\
    rs17585580 & -0.603 & -0.491(0.159) & 0 & -0.153(0.070) & -0.197(0.118) & -0.648(0.136)\\
    Hypertension & 0.368 & 0.324(0.211) & 0.228(0.131) & 0.224(0.133) & 0.246(0.161) & 0.409(0.254)\\
    Diabetes & 1.180 & 1.136(0.216) & 0.769(0.137) & 0.712(0.137) & 0.855(0.161) & 1.260(0.299)\\
    Hypercholesterolomia & 1.403 & 1.294(0.218) & 0.867(0.128) & 0.879(0.132) & 0.992(0.158) & 1.483(0.258)\\
    Sex & 0.361 & 0.325(0.205) & 0.149(0.123) & 0.153(0.125) & 0.205(0.151) & 0.392(0.229)\\
    Smoking & 0.860 & 0.798(0.240) & 0.431(0.145) & 0.431(0.143) & 0.546(0.138) & 0.938(0.235)\\
    HXMI & 37.388 & 42.379(1.097) & 38.919(1.716) & 11.130(0.310) & 12.368(0.435) & 26.677(0.803)\\
    Race(African) & -0.030 & 0.031(0.436) & 0.308(0.283) & 0.114(0.263) & 0.116(0.322) & -0.050(0.474)\\
    Race(Caucasian) & 0.385 & 0.357(0.378) & 0.016(0.225) & 0.123(0.223) & 0.236(0.274) & 0.412(0.411)\\
    \hline
    \end{tabular}}
\end{table}

\subsection*{Scenario 4: Poisson partly linear model}

In this scenario, we present the details of a simulation study for the Poisson partly linear model. Let $q_w = 5$ and $q_z = 4$. The true values of $\boldsymbol{\beta}$ is $\boldsymbol{\beta}_0 = \{1,-0.75,0,\ldots,0,-1,0.75,-0.75\}^\top$, and the true values of $\boldsymbol{\alpha}$ is $\boldsymbol{\alpha}_0 = \{ 0.75,-0.5,-0.5,0.75,-1\}^\top$. We generate $\textbf{X}, \textbf{W}$, and $\textbf{Z}$ from the same distributions used in Scenarios 1 and 2. By setting $\psi_1(z_{i1}) = 0.1(z_{i1}-3)^2$, $\psi_2(z_{i2}) = 0.2(\cos(2\pi z_{i2})+1)$, $\psi_3(z_{i3}) = 0.2\sin(2\pi z_{i3})$, and $\psi_4 (z_{i4}) = 0.1(z_{i4}+1)^3$, we generate $y_i$ from a Poisson distribution with a rate $\lambda_i = \exp \{\boldsymbol{\beta}^\top \textbf{x}_i + \boldsymbol{\alpha}^\top \textbf{w}_i + \sum^4_{j=1} \psi_j(z_{ij}) \}$. From Table \ref{Sce3}, the selection accuracy of GPLM-BAR method is better than the LASSO and Adaptive LASSO methods. We set the number of basis $m_j = 3$, for $j=1,2,3,4$. We also report the values of the individual estimates of $\boldsymbol{\alpha}$ and $\boldsymbol{\beta}$ in Tables \ref{EstalphaPoi} and \ref{EstbetaPoi}, respectively, where the individual BAR estimates are also the least biased among all methods.

\begin{table}
\centering
\caption{Variable selection results over 200 replications for Scenario 3. Standard deviations are in parentheses.} \label{Sce3}
\begin{tabular}{l|llllll}
\hline
 Method & MMSE & TP & FP & MS & MC & TM \\
\hline
\multicolumn{7}{c}{$n=600,p=300$} \\
\hline
BAR(AIC) & 0.003(0.003) & 5 & 0.90 & 5.90 & 0.90 & 46$\%$ \\
BAR(BIC) & 0.002(0.001) & 5 & 0 & 5 & 0 & 100$\%$ \\
LASSO & 0.368(0.967) & 4 & 0.67 & 4.67 & 1.67 & 37$\%$  \\
ALASSO & 0.106(0.418) & 4.87 & 0 & 4.87 & 0.13 & 96$\%$  \\
Oracle & 0.002(0.001) & 5 & 0 & 5 & 0 & 100$\%$ \\
\hline
\multicolumn{7}{c}{$n=800,p=300$} \\
\hline
BAR(AIC) & 0.003(0.002) & 5 & 1.09 & 6.09 & 1.09 & 38$\%$ \\
BAR(BIC) & 0.001(0.001) & 5 & 0.01 & 5.01 & 0.01 & 99.5$\%$ \\
LASSO & 0.365(0.572) & 4.7 & 0.34 & 5.04 & 0.64 & 66$\%$  \\
ALASSO & 0.407(0.320) & 4.96 & 0 & 4.96 & 0.04 & 98$\%$  \\
Oracle & 0.001(0.001) & 5 & 0 & 5 & 0 & 100$\%$ \\
\hline
\multicolumn{7}{c}{$n=600,p=450$} \\
\hline
BAR(AIC) & 0.004(0.004) & 5 & 0.87 & 5.87 & 0.87 & 44$\%$  \\
BAR(BIC) & 0.002(0.002) & 5 & 0 & 5 & 0 & 100$\%$ \\
LASSO & 0.321(1.064) & 3.75 & 0.90 & 4.65 & 2.15 & 28$\%$  \\
ALASSO & 0.100 (0.525) & 4.80 & 0 & 4.80 & 0.20 & 96$\%$  \\
Oracle & 0.002(0.002) & 5 & 0 & 5 & 0 & 100$\%$ \\
\hline
\multicolumn{7}{c}{$n=800,p=450$} \\
\hline
BAR(AIC) & 0.003(0.003) & 5 & 1.18 & 6.18 & 1.18 & 34$\%$  \\
BAR(BIC) & 0.001(0.001) & 5 & 0 & 5 & 0 & 100$\%$ \\
LASSO & 0.310(1.101) & 3.65 & 0.38 & 4.03 & 1.73 & 47$\%$  \\
ALASSO & 0.102(0.305) & 4.95 & 0 & 4.95 & 0.05 & 98$\%$  \\ 
Oracle & 0.001(0.001) & 5 & 0 & 5 & 0 & 100$\%$ \\
\hline
\end{tabular}
\end{table}

\begin{table}
\centering
\caption{Estimation results of non-zero entries in $\boldsymbol{\beta}$ for Scenario 3. Standard deviations are in parentheses.} \label{EstbetaPoi}
\begin{tabular}{l | lllll}
\hline
Method & Bias($\widehat{\beta}_1$) & Bias($\widehat{\beta}_2$) & Bias($\widehat{\beta}_{p-2}$) & Bias($\widehat{\beta}_{p-1}$) & Bias($\widehat{\beta}_{p}$)  \\
\hline
\multicolumn{6}{c}{$n=600,p=300$} \\
\hline
BAR(AIC) & -0.001(0.021) & 0.002(0.021) & 0.003(0.024) & -0.001(0.023) & 0.002(0.022) \\
BAR(BIC) & -0.005(0.020) & 0.006(0.021) & 0.007(0.023) & -0.006(0.022) & 0.006(0.022) \\
LASSO & -0.440(0.286) & 0.385(0.193) & 0.438(0.285) & -0.425(0.177) & 0.374(0.200) \\
ALASSO & -0.179(0.141) & 0.204(0.130) & 0.189(0.148) & -0.257(0.158) & 0.206(0.138) \\
Oracle & 0(0.021) & 0(0.020) & 0.002(0.023) & 0.001(0.022) & 0(0.022) \\
\hline
\multicolumn{6}{c}{$n=800,p=300$} \\
\hline
BAR(AIC) & 0.0006(0.018) & -0.002(0.019) & 0.002(0.02) & -0.001(0.018) & 0.002(0.017) \\
BAR(BIC) & -0.002(0.018) & 0.002(0.019) & 0.005(0.02) & -0.004(0.017) & 0.005(0.017) \\
LASSO & -0.355(0.173) & 0.327(0.125) & 0.349(0.171) & -0.378(0.118)  & 0.331(0.124) \\
ALASSO & -0.244(0.127) & 0.278(0.148) & 0.245(0.125) & -0.349(0.179) & 0.291(0.151) \\
Oracle & 0.001(0.018)& -0.002(0.019) & 0.001(0.020)& 0.001(0.017)& 0.001(0.017) \\
\hline
\multicolumn{6}{c}{$n=600,p=450$} \\
\hline
BAR(AIC) & 0.003(0.022) & 0.001(0.020) & 0.003(0.025) & 0.001(0.025) & 0.003(0.021) \\
BAR(BIC) & -0.001(0.022) & 0.004(0.020) & 0.006(0.024) & -0.004(0.024) & 0.008(0.021) \\
LASSO & -0.459(0.318) & 0.399(0.213) & 0.454(0.320) & -0.43(0.197) & 0.396(0.213) \\
ALASSO & -0.177(0.175) & 0.195(0.128) & 0.179(0.175) & -0.238(0.136) & 0.198(0.132) \\
Oracle & 0.004(0.022) & -0.001(0.020) & 0.001(0.024) & 0.003(0.024) & 0.002(0.021) \\
\hline
\multicolumn{6}{c}{$n=800,p=450$} \\
\hline
BAR(AIC) & 0.001(0.020) & -0.001(0.017) & -0.001(0.019) & -0.001(0.019) & -0.001(0.019) \\
BAR(BIC) & -0.001(0.020) & 0.002(0.017) & 0.001(0.018) & -0.004(0.019) & 0.002(0.019) \\
LASSO & -0.469(0.327) & 0.393(0.224) & 0.462(0.330) & -0.430(0.203) & 0.391(0.224) \\
ALASSO & -0.186(0.117) & 0.215(0.134) & 0.188(0.117) & -0.284(0.166) & 0.220(0.133) \\
Oracle & 0.002(0.020) & -0.002(0.017) & -0.002(0.018) & 0.001(0.019) & -0.002(0.019) \\
\hline
\end{tabular}
\end{table}

\begin{table}
\centering
\caption{Estimation results of $\boldsymbol{\alpha}$ for Scenario 3. Standard deviations are in parentheses.} \label{EstalphaPoi}
\begin{tabular}{l | lllll}
\hline
Method & Bias($\widehat{\alpha}_1$) & Bias($\widehat{\alpha}_2$) & Bias($\widehat{\alpha}_3$) & Bias($\widehat{\alpha}_4$) & Bias($\widehat{\alpha}_5$)  \\
\hline
\multicolumn{6}{c}{$n=600,p=300$} \\
\hline
BAR(AIC) & -0.001(0.021) & 0.002(0.021) & 0.003(0.024) & -0.001(0.023) & 0.002(0.022) \\
BAR(BIC) & -0.005(0.020) & 0.006(0.021) & 0.007(0.023) & -0.006(0.022) & 0.006(0.022) \\
LASSO & -0.440(0.286) & 0.385(0.193) & 0.438(0.285) & -0.425(0.177) & 0.374(0.200) \\
ALASSO & -0.179(0.141) & 0.204(0.130) & 0.189(0.148) & -0.257(0.158) & 0.206(0.138) \\
Oracle & 0(0.021) & 0(0.020) & 0.002(0.023) & 0.001(0.022) & 0(0.022) \\
\hline
\multicolumn{6}{c}{$n=800,p=300$} \\
\hline
BAR(AIC) & 0.0004(0.037) & 0.002(0.037) & -0.001(0.035) & 0.004(0.034) & -0.001(0.034) \\
BAR(BIC) & 0.0003(0.037) & 0.002(0.036) & 0(0.035) & 0.004(0.033) & -0.0002(0.035) \\
LASSO & -0.085(0.204) & 0.067(0.157) & 0.056(0.164) & -0.080(0.204)  & 0.107(0.253) \\
ALASSO & -0.032(0.134) & 0.042(0.135) & 0.016(0.125) & -0.033(0.136) & 0.035(0.133) \\
Oracle & 0.001(0.037) & 0.002(0.036) & -0.0003(0.035) & 0.005(0.032) & -0.001(0.035)\\
\hline
\multicolumn{6}{c}{$n=600,p=450$} \\
\hline
BAR(AIC) & 0.003(0.044) & -0.001(0.040) & 0.005(0.043) & -0.004(0.045) & 0.002(0.044) \\
BAR(BIC) & 0.0003(0.044) & 0.001(0.039) & 0.006(0.042) & -0.003(0.045) & 0.003(0.043) \\
LASSO & -0.242(0.314) & 0.152(0.220) & 0.146(0.228) & -0.228(0.319) & 0.291(0.424) \\
ALASSO & -0.066(0.161) & 0.040(0.119) & 0.037(0.123) & -0.065(0.161) & 0.071(0.208) \\
Oracle & 0.002(0.044) & 0.0004(0.040) & 0.005(0.042) & -0.002(0.045) & 0.002(0.043)\\
\hline
\multicolumn{6}{c}{$n=800,p=450$} \\
\hline
BAR(AIC) & 0.002(0.035) & -0.005(0.034) & 0.0003(0.035) & 0.001(0.036) & -0.006(0.038) \\
BAR(BIC) & 0.002(0.034) & -0.005(0.034) & 0.001(0.034) & 0.001(0.036) & -0.005(0.038) \\
LASSO & -0.239(0.322) & 0.152(0.229) & 0.159(0.223) & -0.228(0.331) & 0.304(0.433) \\
ALASSO & -0.021(0.119) & 0.018(0.104) & 0.018(0.115) & -0.011(0.123) & 0.023(0.124) \\
Oracle & 0.003(0.034) & -0.005(0.034)& 0.001(0.034)& 0.002(0.036)& -0.006(0.038)\\
\hline
\end{tabular}
\end{table}

\begin{figure}
    \centering
    \includegraphics[height = 12cm, width = 16cm]{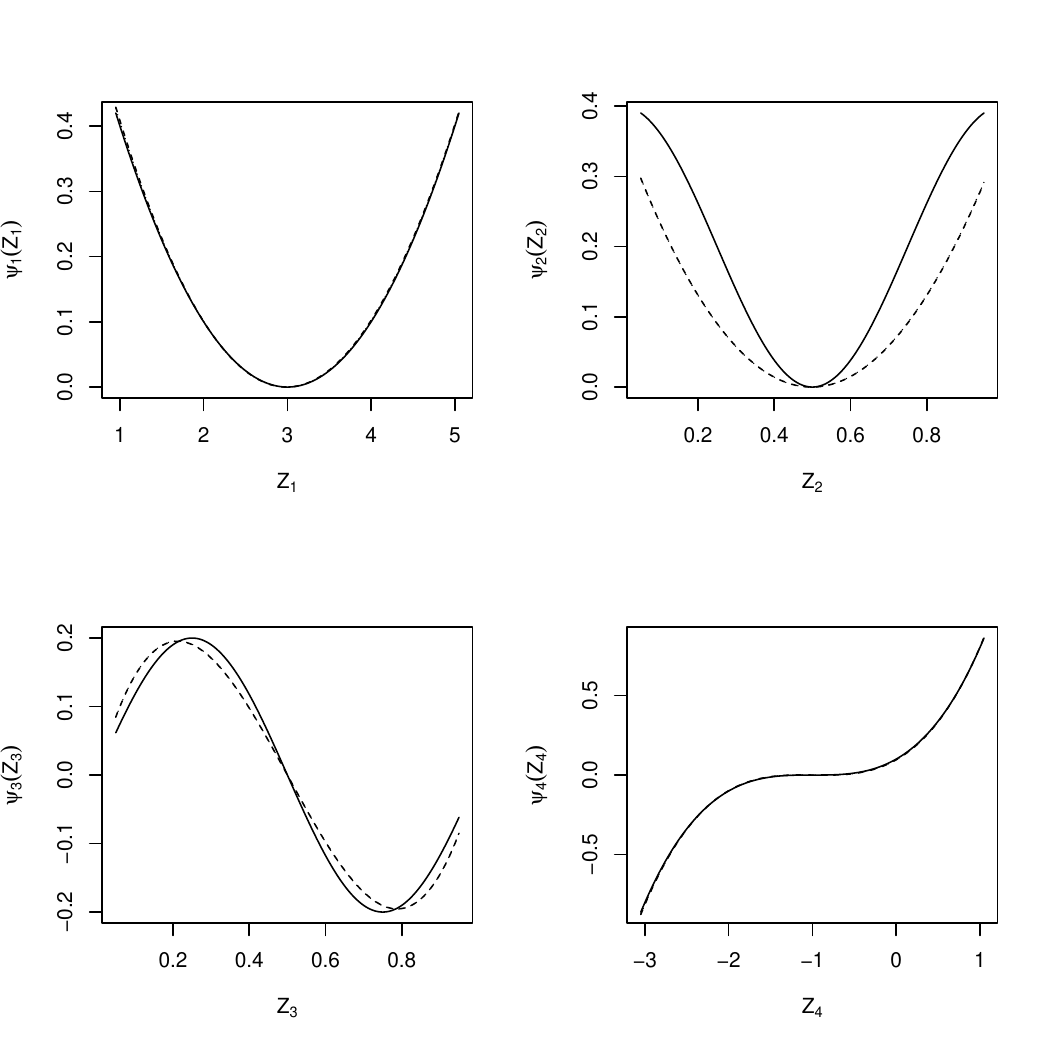}
    \caption{Estimated nonlinear covariate effects $\psi_j, j=1,2,3,4$ for $p=300$ and $n=600$ for Scenario 3. The dashed line represents the GPLM-BAR with AIC penalty, and the dotted line represents GPLM-BAR method with BIC penalty.}
\end{figure}

\begin{figure}
    \centering
    \includegraphics[height = 12cm, width = 16cm]{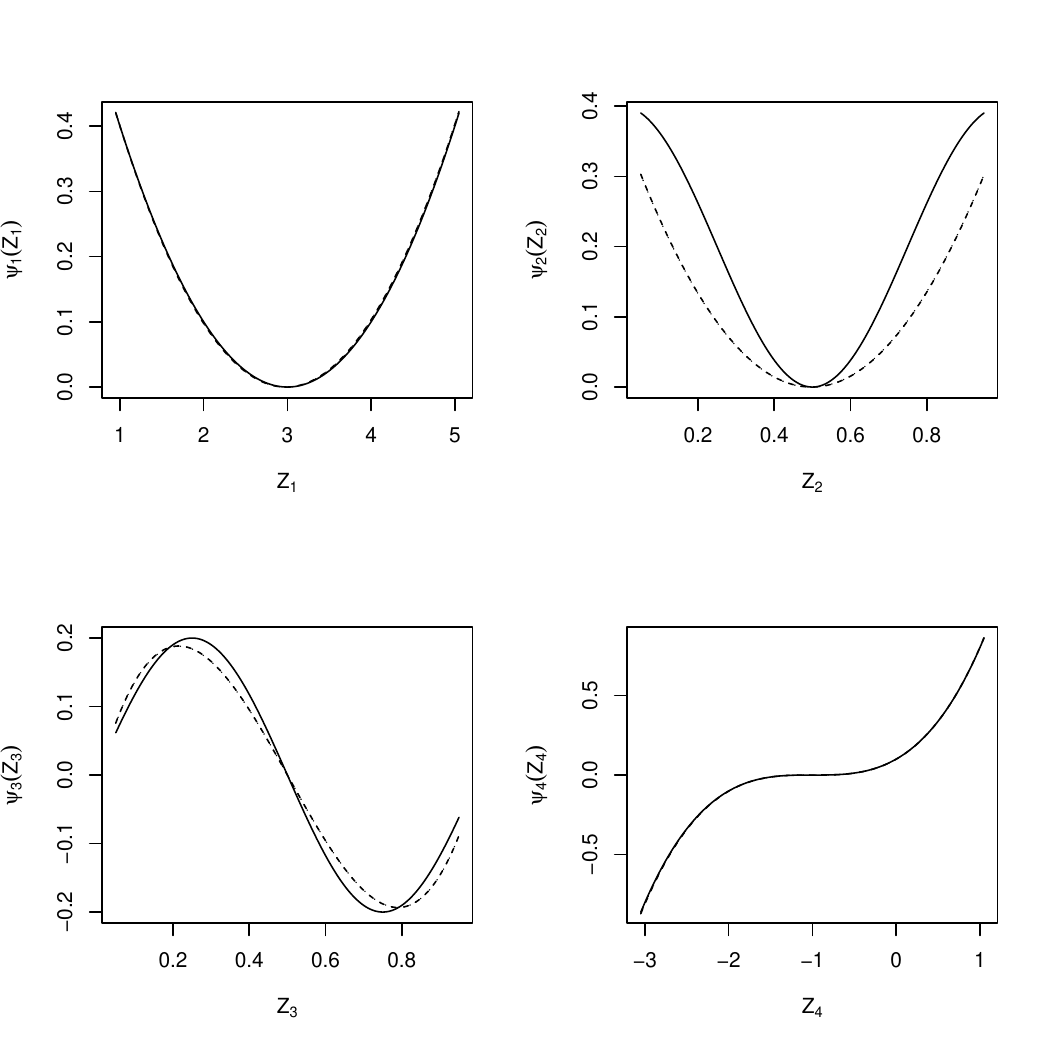}
    \caption{Estimated nonlinear covariate effects $\psi_j, j=1,2,3,4$ for $p=300$ and $n=800$ for Scenario 3. The dashed line represents the GPLM-BAR with AIC penalty, and the dotted line represents GPLM-BAR method with BIC penalty.}
\end{figure}

\begin{figure}
    \centering
    \includegraphics[height = 12cm, width = 16cm]{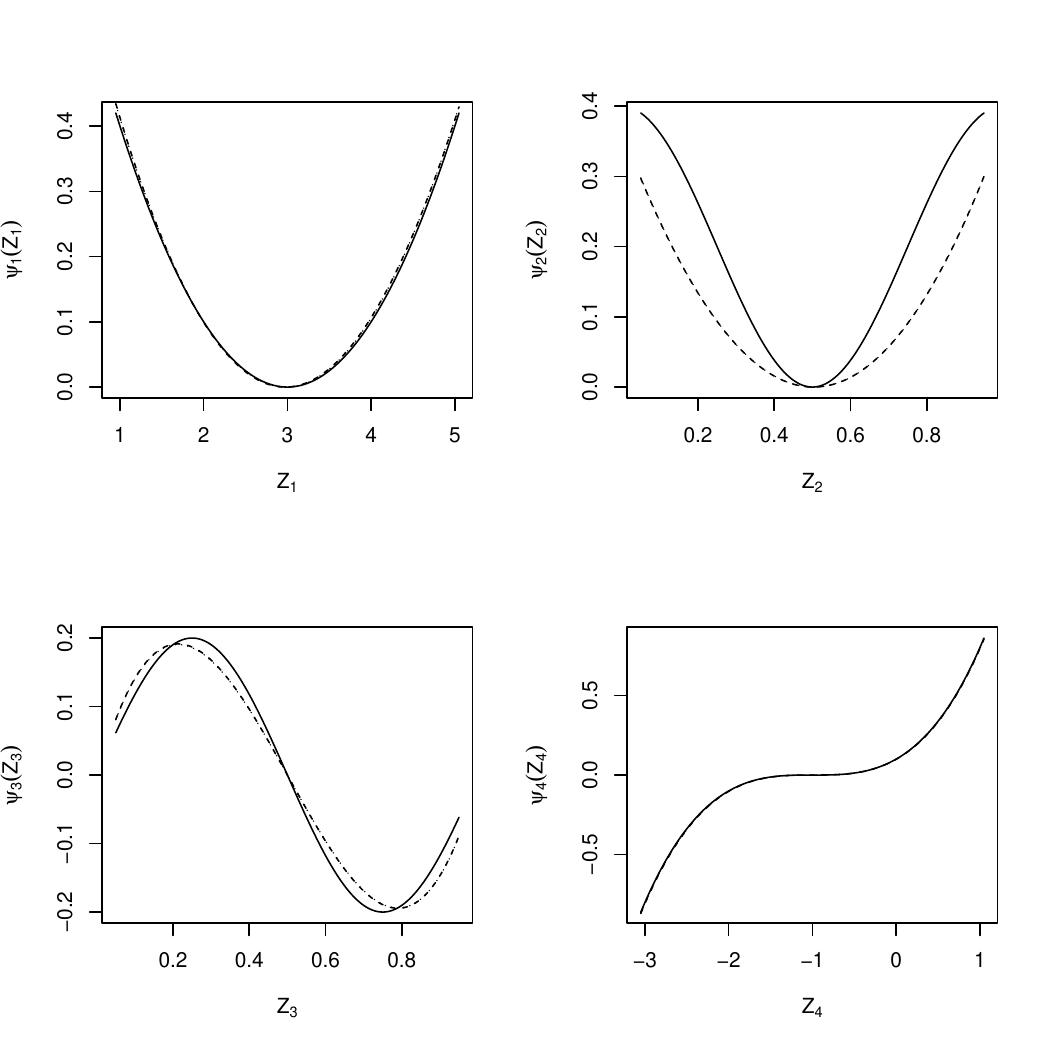}
    \caption{Estimated nonlinear covariate effects $\psi_j, j=1,2,3,4$ for $p=450$ and $n=800$ for Scenario 3. The dashed line represents the GPLM-BAR with AIC penalty, and the dotted line represents GPLM-BAR method with BIC penalty.}
\end{figure}

\subsection*{More simulation results for the logistic partly linear model}

In this section, we present more simulation results that could not be included in the main article, by presenting the following tables, where Scenarios 1 and 2 are for the logistic partly linear model and Scenario 3 is for the Poisson partly linear model. 

\begin{table}
\centering
\caption{Estimation results of non-zero elements in $\boldsymbol{\beta}$ in Scenario 1. Standard deviations are in parentheses.} \label{BetasS1}
\begin{tabular}{l | lllll}
\hline
Method & Bias($\widehat{\beta}_1$) & Bias($\widehat{\beta}_2$) & Bias($\widehat{\beta}_{p-2}$) & Bias($\widehat{\beta}_{p-1}$) & Bias($\widehat{\beta}_{p}$)  \\
\hline
\multicolumn{6}{c}{$n=600,p=300$} \\
\hline
BAR(AIC) & -0.01(0.15) & 0.005(0.14) & 0.01(0.13) & -0.03(0.14) & -0.01(0.14) \\
BAR(BIC) & -0.24(0.15) & 0.23(0.15) & 0.26(0.13) & -0.27(0.24) & -0.19(0.22) \\
LASSO & -0.49(0.11) & 0.48(0.11) & 0.49(0.10) & -0.40(0.10) & -0.32(0.11) \\
ALASSO & -0.23(0.16) & 0.22(0.16) & 0.23(0.15) & -0.22(0.16) & -0.14(0.15) \\
Oracle & 0.05(0.15) & -0.06(0.14) & -0.06(0.13) & 0.04(0.13) & 0.04(0.14) \\
\hline
\multicolumn{6}{c}{$n=800,p=300$} \\
\hline
BAR(AIC) & -0.01(0.11) & 0.01(0.11) & 0.01(0.11) & -0.02(0.12) & -0.02(0.12) \\
BAR(BIC) & -0.18(0.11) & 0.17(0.11) & 0.18(0.11) & -0.17(0.15) & -0.15(0.14) \\
LASSO & -0.43(0.09) & 0.43(0.09) & 0.43(0.09) & -0.35(0.09)  & -0.29(0.09) \\
ALASSO & -0.20(0.13) & 0.19(0.13) & 0.20(0.12) & -0.18(0.12) & -0.14(0.12) \\
Oracle & 0.04(0.12) & -0.04(0.12) & -0.04(0.11) & 0.03(0.12) & 0.02(0.11) \\
\hline
\multicolumn{6}{c}{$n=600,p=450$} \\
\hline
BAR(AIC) & -0.01(0.15) & 0.004(0.16) & 0.01(0.14) & -0.03(0.14) & -0.01(0.13) \\
BAR(BIC) & -0.23(0.15) & 0.23(0.17) & 0.27(0.17) & -0.29(0.25) & -0.19(0.20) \\
LASSO & -0.50(0.11) & 0.49(0.12) & 0.50(0.10) & -0.41(0.10) & -0.33(0.09) \\
ALASSO & -0.22(0.15) & 0.22(0.16) & 0.22(0.14) & -0.21(0.15) & -0.14(0.13) \\
Oracle & 0.06(0.15) & -0.06(0.16) & -0.06(0.14) & 0.03(0.13) & 0.04(0.13) \\
\hline
\multicolumn{6}{c}{$n=800,p=450$} \\
\hline
BAR(AIC) & -0.01(0.21) & 0.01(0.21) & -0.001(0.18) & -0.001(0.18) & -0.01(0.20) \\
BAR(BIC) & -0.17(0.19) & 0.18(0.19) & 0.17(0.16) & -0.16(0.17) & -0.14(0.18) \\
LASSO & -0.44(0.17) & 0.45(0.17) & 0.44(0.15) & -0.35(0.15) & -0.30(0.16) \\
ALASSO & -0.19(0.20) & 0.19(0.19) & 0.18(0.17) & -0.15(0.17) & -0.13(0.19) \\
Oracle & 0.04(0.12) & -0.04(0.13) & -0.05(0.13) & 0.04(0.12) & 0.02(0.11) \\
\hline
\end{tabular}
\end{table}

\begin{table}
\centering
\caption{Estimation results of non-zero elements in $\boldsymbol{\beta}$ for Scenario 2. Standard deviations are in parentheses.} \label{BetasS2}
\begin{tabular}{l | lllll}
\hline
Method & Bias($\widehat{\beta}_1$) & Bias($\widehat{\beta}_2$) & Bias($\widehat{\beta}_{p-2}$) & Bias($\widehat{\beta}_{p-1}$) & Bias($\widehat{\beta}_{p}$)  \\
\hline
\multicolumn{6}{c}{$n=600,p=300$} \\
\hline
BAR(AIC) & -0.08(0.16) & 0.08(0.15) & 0.09(0.15) & -0.11(0.18) & -0.07(0.18) \\
BAR(BIC) & -0.46(0.12) & 0.46(0.11) & 0.46(0.12) & -0.35(0.09) & -0.30(0.15) \\
LASSO & -0.33(0.11) & 0.34(0.10) & 0.33(0.10) & -0.28(0.09) & -0.21(0.10) \\
ALASSO & -0.22(0.15) & 0.22(0.14) & 0.21(0.14) & -0.21(0.13) & -0.12(0.13) \\
Oracle & 0.02(0.11) & -0.02(0.10) & -0.02(0.11) & 0.01(0.11) & 0.01(0.12) \\
\hline
\multicolumn{6}{c}{$n=800,p=300$} \\
\hline
BAR(AIC) & -0.06(0.11) & 0.05(0.11) & 0.06(0.11) & -0.07(0.15) & -0.04(0.13) \\
BAR(BIC) & -0.41(0.16) & 0.40(0.17) & 0.41(0.16) & -0.32(0.13) & -0.26(0.18) \\
LASSO & -0.30(0.09) & 0.29(0.09) & 0.29(0.08) & -0.24(0.09) & -0.18(0.08) \\
ALASSO & -0.19(0.12) & 0.18(0.12) & 0.18(0.12) & -0.17(0.12) & -0.10(0.11) \\
Oracle & 0.004(0.09) & -0.01(0.10) & -0.01(0.09) & 0.01(0.10) & 0.01(0.10) \\
\hline
\multicolumn{6}{c}{$n=600,p=450$} \\
\hline
BAR(AIC) & -0.06(0.15) & 0.06(0.14) & 0.10(0.15) & -0.10(0.19) & -0.06(0.17) \\
BAR(BIC) & -0.45(0.13) & 0.45(0.13) & 0.48(0.09) & -0.36(0.08) & -0.31(0.15) \\
LASSO & -0.34(0.10) & 0.34(0.10) & 0.35(0.09) & -0.28(0.08) & -0.22(0.10) \\
ALASSO & -0.20(0.15) & 0.20(0.14) & 0.22(0.13) & -0.19(0.14) & -0.12(0.14) \\
Oracle & 0.03(0.12) & -0.03(0.11) & -0.01(0.11) & 0.02(0.12) & 0.02(0.11) \\
\hline
\multicolumn{6}{c}{$n=800,p=450$} \\
\hline
BAR(AIC) & -0.04(0.11) & 0.04(0.11) & 0.05(0.12) & -0.05(0.14) & -0.05(0.15) \\
BAR(BIC) & -0.38(0.18) & 0.38(0.18) & 0.39(0.17) & -0.31(0.14) & -0.24(0.18) \\
LASSO & -0.30(0.09) & 0.30(0.09) & 0.30(0.09) & -0.24(0.09) & -0.19(0.09) \\
ALASSO & -0.16(0.12) & 0.16(0.12) & 0.17(0.12) & -0.15(0.13) & -0.10(0.13) \\
Oracle & 0.02(0.10) & -0.02(0.10) & -0.02(0.10) & 0.02(0.10) & 0.01(0.10) \\
\hline
\end{tabular}
\end{table}

\begin{table}
\centering
\caption{Estimation results of $\boldsymbol{\alpha}$ for Scenario 1. Standard deviations are in parentheses.}\label{alphaS1}
\begin{tabular}{l|lllll}
\hline
Method & $\text{Bias}(\widehat{\alpha}_1)$ & $\text{Bias}(\widehat{\alpha}_2)$ & $\text{Bias}(\widehat{\alpha}_3)$ & $\text{Bias}(\widehat{\alpha}_4)$ & $\text{Bias}(\widehat{\alpha}_5)$ \\
\hline
\multicolumn{6}{c}{$n=600,p=300$} \\
\hline
BAR(AIC) & 0.02(0.26) & -0.04(0.23) & 0.01(0.22) & 0.01(0.22) & -0.001(0.24) \\
BAR(BIC) & -0.10(0.23) & -0.03(0.20) & 0.07(0.19) & -0.09(0.20) & 0.13(0.21) \\
LASSO & -0.21(0.20) & 0.09(0.18) & 0.12(0.17) & -0.16(0.18) & 0.23(0.19)\\
ALASSO & -0.09(0.22) & 0.02(0.20) & 0.07(0.19) & -0.08(0.20) & 0.12(0.22)\\
Oracle & 0.06(0.26)& -0.06(0.24) & 0(0.22) & 0.03(0.23) & -0.03(0.24) \\
\hline
\multicolumn{6}{c}{$n=800,p=300$} \\
\hline
BAR(AIC) & 0.01(0.20) & -0.02(0.19) & -0.005(0.19)& 0.03(0.19) & -0.02(0.21)\\
BAR(BIC) & -0.09(0.18) & 0.04(0.17) & 0.04(0.17) & -0.05(0.17) & 0.07(0.19)\\
LASSO & -0.20(0.16) & 0.09(0.15) & 0.10(0.15) & -0.14(0.15) & 0.18(0.17)\\
ALASSO & -0.09(0.18) & 0.04(0.17) & 0.05(0.17) & -0.05(0.17) & 0.08(0.20)\\
Oracle & 0.04(0.20) & -0.02(0.20) & -0.01(0.19) & 0.04(0.20) & 0.05(0.21)\\
\hline
\multicolumn{6}{c}{$n=600,p=450$} \\
\hline
BAR(AIC) & 0.06(0.23) & -0.02(0.22) & -0.01(0.25) & 0.02(0.24)& -0.07(0.24)\\
BAR(BIC) & -0.08(0.20) & 0.06(0.19)& 0.06(0.21) & -0.08(0.21) & 0.07(0.21) \\
LASSO & -0.19(0.18) & 0.11(0.17) & 0.11(0.18) & -0.16(0.19) & 0.18(0.19)\\
ALASSO & -0.06(0.21) & 0.04(0.19) & 0.05(0.21) & -0.06(0.22) & 0.05(0.22)\\
Oracle & 0.09(0.23) & -0.02(0.22) & -0.02(0.25) & 0.05(0.24) & 0.10(0.25) \\
\hline
\multicolumn{6}{c}{$n=800,p=450$} \\
\hline
BAR(AIC) & 0.02(0.21) & -0.01(0.21) & -0.005(0.18)& 0.03(0.18) & -0.03(0.20)\\
BAR(BIC) & -0.08(0.19) & 0.04(0.19) & 0.05(0.16) & -0.05(0.17) & 0.08(0.18)\\
LASSO & -0.20(0.17) & 0.10(0.17) & 0.11(0.15) & -0.14(0.15) & 0.19(0.16)\\
ALASSO & -0.08(0.20) & 0.04(0.19) & 0.05(0.17) & -0.05(0.17) & 0.07(0.19)\\
Oracle & 0.04(0.21) & -0.02(0.21) & -0.01(0.18) & 0.05(0.19) & -0.04(0.20) \\
\hline
\end{tabular}
\end{table}

\begin{table}
\centering
\caption{Estimation results of $\boldsymbol{\alpha}$ for Scenario 2. Standard deviations are in parentheses.}\label{alphaS2}
\begin{tabular}{l|lllll}
\hline
Method & $\text{Bias}(\widehat{\alpha}_1)$ & $\text{Bias}(\widehat{\alpha}_2)$ & $\text{Bias}(\widehat{\alpha}_3)$ & $\text{Bias}(\widehat{\alpha}_4)$ & $\text{Bias}(\widehat{\alpha}_5)$ \\
\hline
\multicolumn{6}{c}{$n=600,p=300$} \\
\hline
BAR(AIC) & 0.04(0.21) & -0.02(0.22) & -0.03(0.20) & 0.005(0.22) & -0.004(0.20) \\
BAR(BIC) & -0.07(0.19) & 0.04(0.20) & 0.03(0.19) & -0.07(0.21) & 0.11(0.18) \\
LASSO & -0.06(0.19) & 0.03(0.20) & 0.02(0.19) & -0.06(0.20) & 0.09(0.18) \\
ALASSO & -0.02(0.19) & 0.01(0.21) & 0.005(0.19) & -0.04(0.21) & 0.05(0.18) \\
Oracle & 0.07(0.21) & -0.04(0.22) & -0.04(0.21) & 0.03(0.22) & 0.04(0.20) \\
\hline
\multicolumn{6}{c}{$n=800,p=300$} \\
\hline
BAR(AIC) & 0.005(0.18) & 0.01(0.17) & -0.02(0.18) & 0.02(0.17) & -0.03(0.17) \\
BAR(BIC) & -0.10(0.17) & 0.06(0.17) & 0.03(0.17) & -0.07(0.16) & 0.07(0.16) \\
LASSO & -0.09(0.17) & 0.05(0.16) & 0.02(0.17) & -0.06(0.15) & 0.06(0.15) \\
ALASSO & -0.05(0.17) & 0.03(0.16) & 0.004(0.17) & -0.03(0.15) & 0.03(0.15) \\
Oracle & 0.03(0.18) & -0.003(0.17) & -0.04(0.18) & 0.03(0.17) & -0.05(0.17)\\
\hline
\multicolumn{6}{c}{$n=600,p=450$} \\
\hline
BAR(AIC) & 0.02(0.21) & -0.01(0.21) & -0.01(0.21) & 0.03(0.19) & -0.05(0.21) \\
BAR(BIC) & -0.11(0.20) & 0.05(0.20) & 0.06(0.20) & -0.07(0.17) & 0.08(0.20) \\
LASSO & -0.10(0.19) & 0.05(0.20) & 0.06(0.19) & -0.07(0.16) & 0.08(0.19) \\
ALASSO & -0.05(0.20) & 0.02(0.20) & 0.03(0.19) & -0.03(0.18) & 0.03(0.20) \\
Oracle & 0.05(0.21) & -0.03(0.21) & -0.02(0.21) & 0.05(0.19) & 0.08(0.21) \\
\hline
\multicolumn{6}{c}{$n=800,p=450$} \\
\hline
BAR(AIC) & 0.02(0.21) & -0.01(0.21) & 0.005(0.18) & 0.03(0.18) & -0.03(0.20) \\
BAR(BIC) & -0.08(0.19) & 0.04(0.19) & 0.05(0.16) & -0.05(0.17) & 0.08(0.18) \\
LASSO & -0.22(0.16) & 0.11(0.16) & 0.12(0.14) & -0.16(0.14) & 0.22(0.16) \\
ALASSO & -0.12(0.20) & 0.06(0.18) & 0.07(0.16) & -0.08(0.16) & 0.12(0.18) \\
Oracle & 0.03(0.19) & -0.01(0.21) & 0.01(0.20) & 0.05(0.18) & 0.04(0.18) \\
\hline
\end{tabular}
\end{table}

\begin{figure}
\centering
\includegraphics[height=12cm, width=16cm]{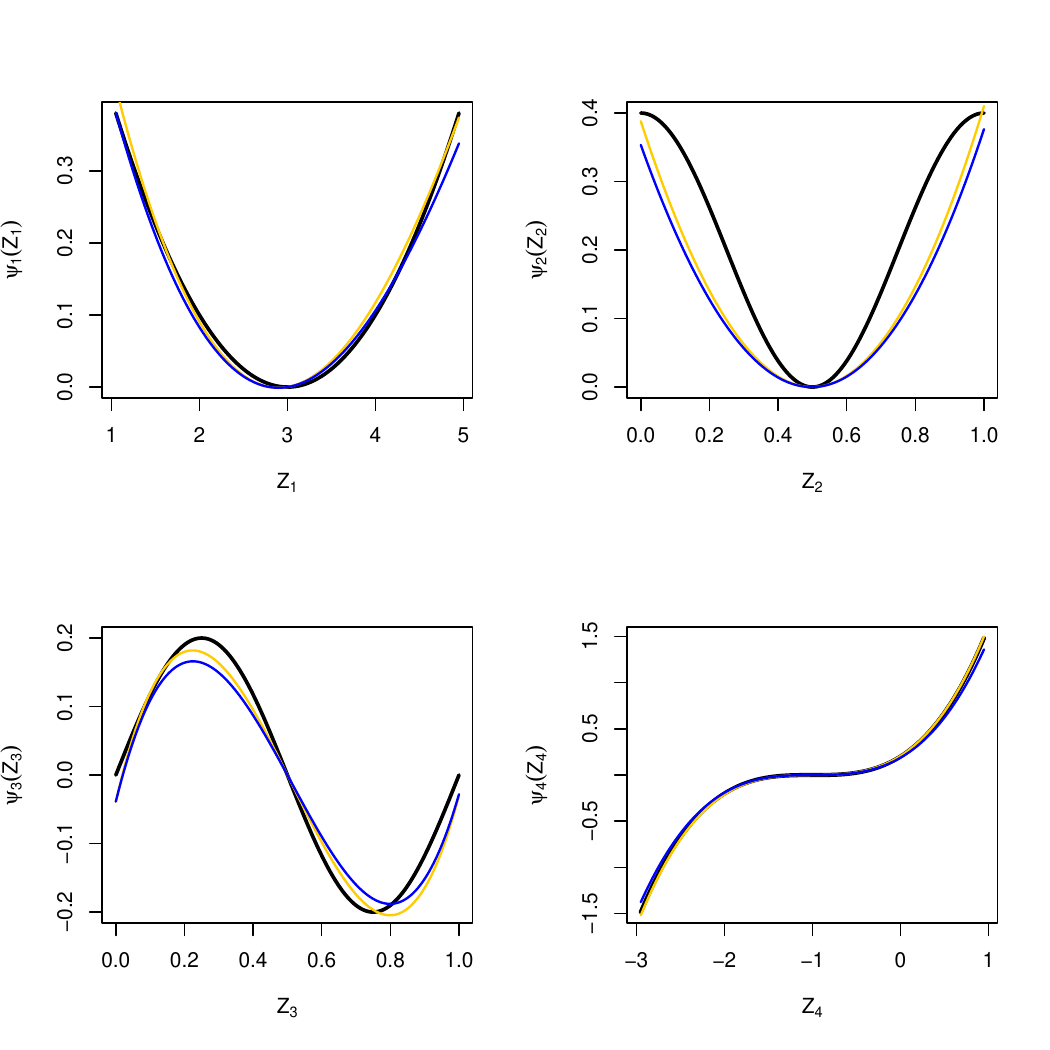}
\caption{Estimated nonlinear covariate effects $\psi_j, j=1,2,3,4$ for $p=300$ and $n=800$ for Scenario 1. The yellow curve represents the GPLM-BAR with AIC penalty, and the blue curve represents GPLM-BAR method with BIC penalty.}\label{ESTS1p300n800}
\end{figure}

\begin{figure}
\centering
\includegraphics[height=12cm, width=16cm]{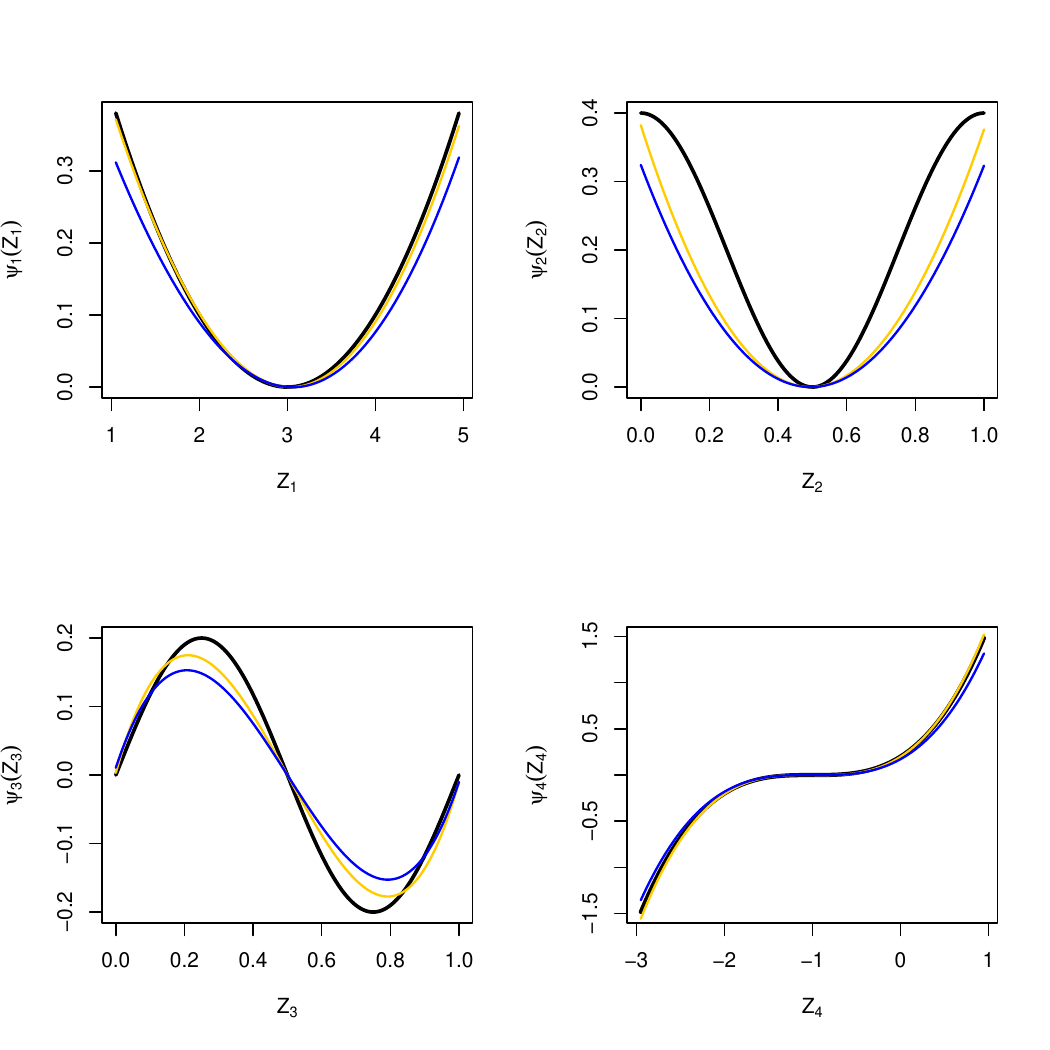}
\caption{Estimated nonlinear covariate effects $\psi_j, j=1,2,3,4$ for $p=450$ and $n=600$ for Scenario 1. The yellow curve represents the GPLM-BAR with AIC penalty, and the blue curve represents GPLM-BAR method with BIC penalty.}\label{ESTS1p450n600}
\end{figure}

\begin{figure}
\centering
\includegraphics[height=12cm, width=16cm]{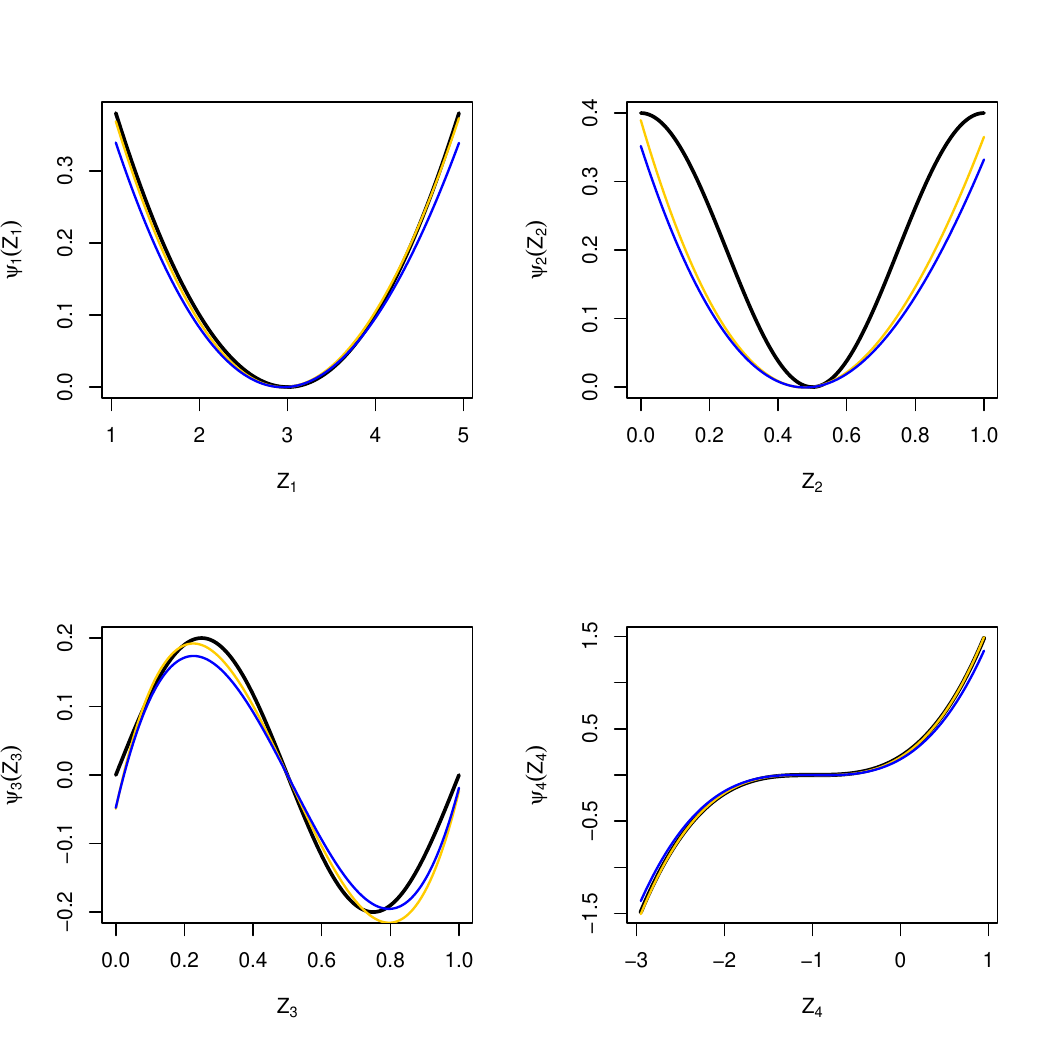}
\caption{Estimated nonlinear covariate effects $\psi_j, j=1,2,3,4$ for $p=450$ and $n=800$ for Scenario 1. The yellow curve represents the GPLM-BAR with AIC penalty, and the blue curve represents GPLM-BAR method with BIC penalty.}\label{ESTS1p450n800}
\end{figure}

\begin{figure}
\centering
\includegraphics[height=12cm, width=16cm]{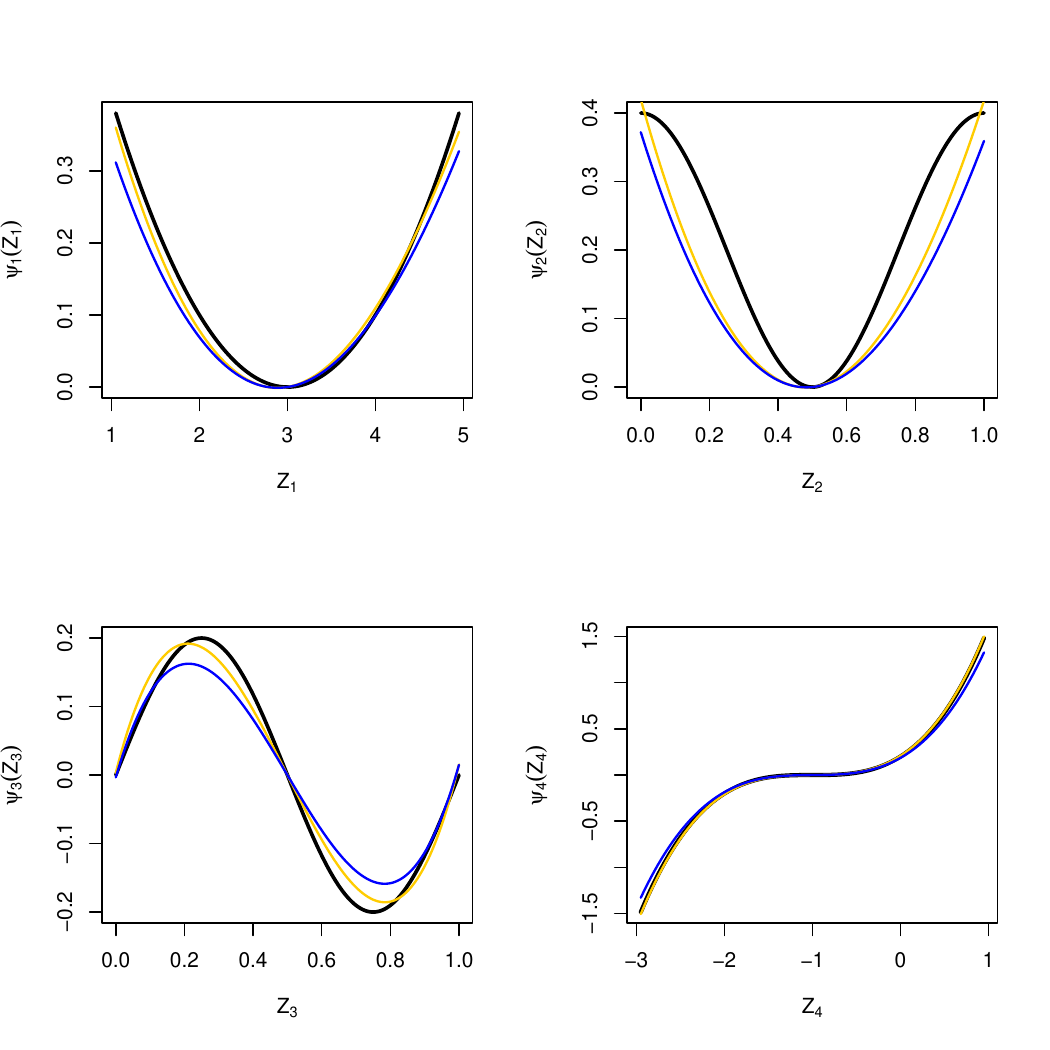}
\caption{Estimated nonlinear covariate effects $\psi_j, j=1,2,3,4$ for $p=300$ and $n=600$ for Scenario 2. The yellow curve represents the GPLM-BAR with AIC penalty, and the blue curve represents GPLM-BAR method with BIC penalty.}\label{ESTS2p300n600}
\end{figure}

\begin{figure} 
\centering
\includegraphics[height=12cm, width=16cm]{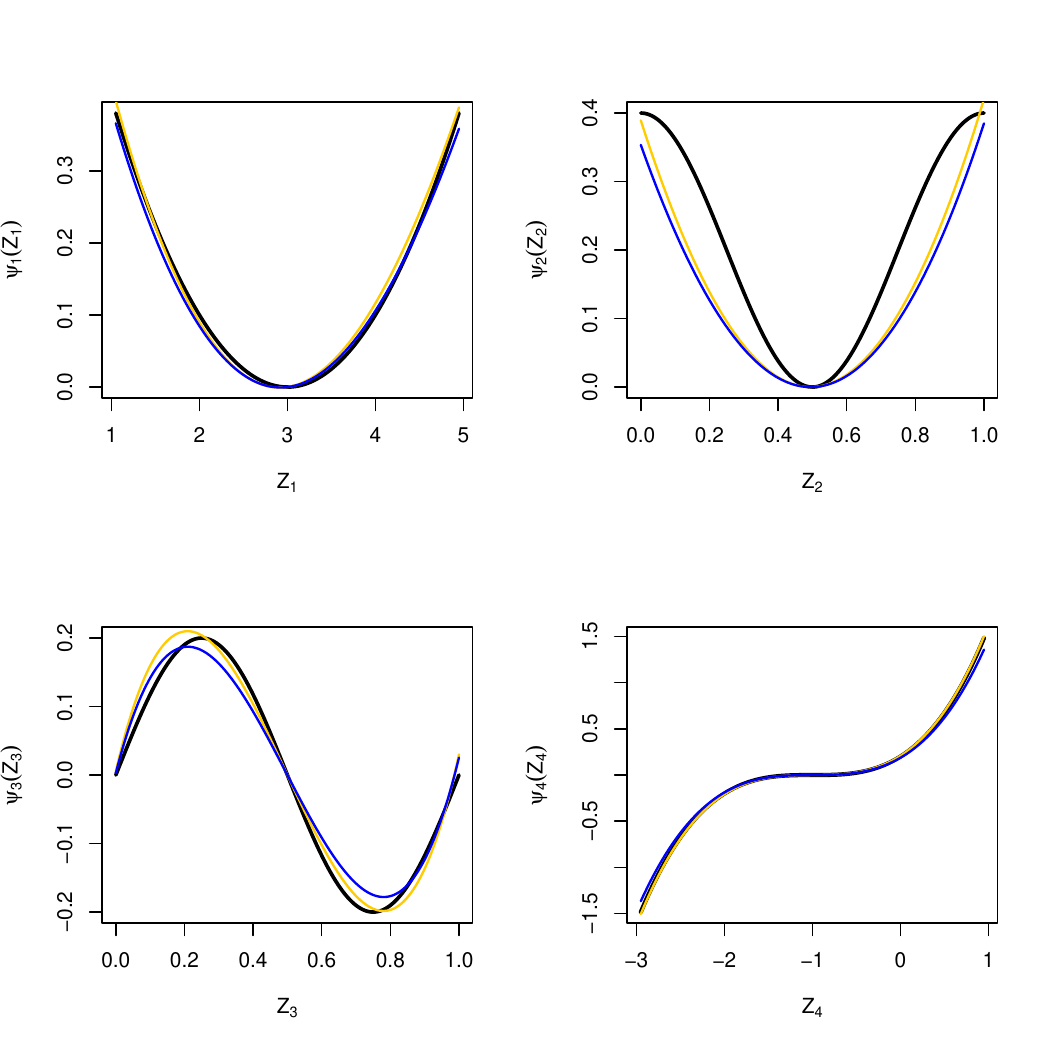}
\caption{Estimated nonlinear covariate effects $\psi_j, j=1,2,3,4$ for $p=300$ and $n=800$ for Scenario 2. The yellow curve represents the GPLM-BAR with AIC penalty, and the blue curve represents GPLM-BAR method with BIC penalty.}\label{ESTS2p300n800}
\end{figure}

\begin{figure}
\centering
\includegraphics[height=12cm, width=16cm]{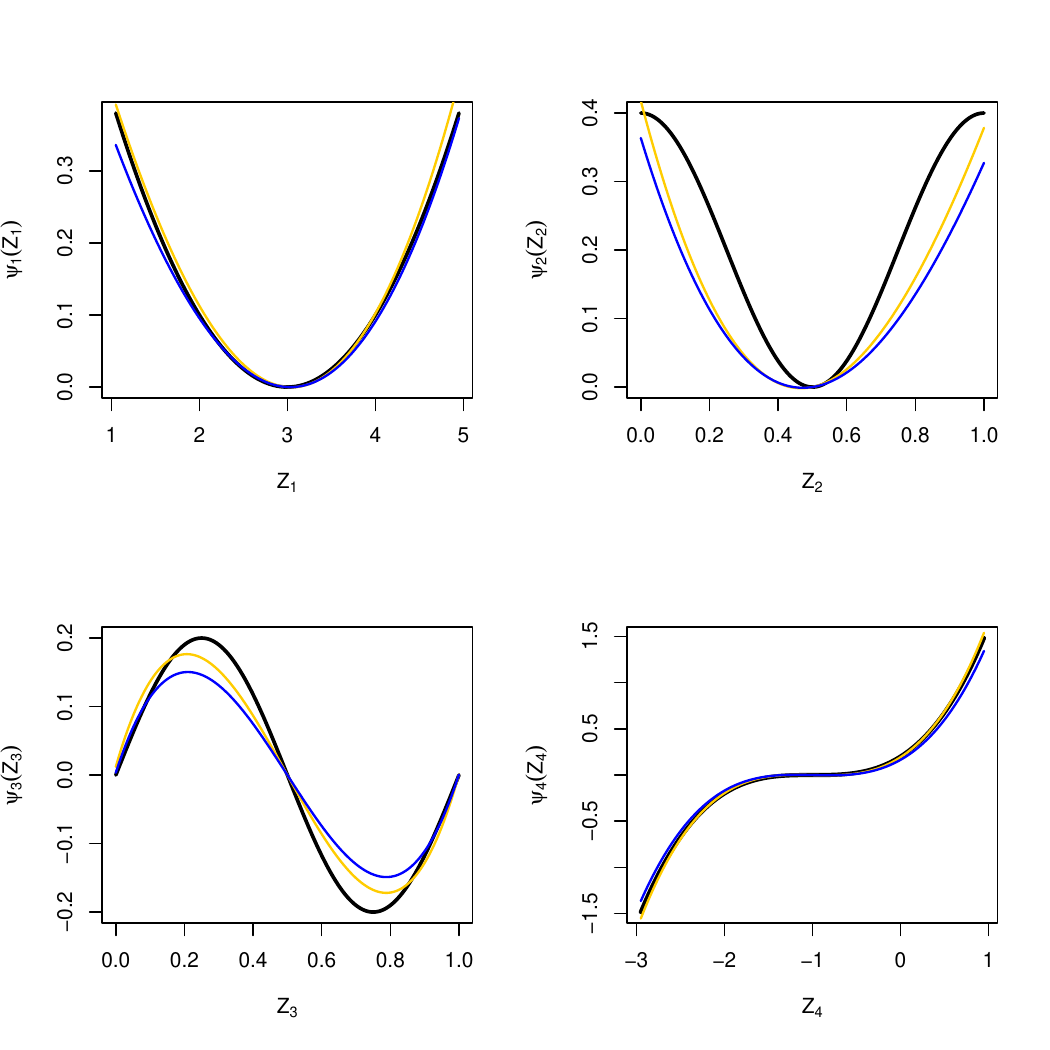}
\caption{Estimated nonlinear covariate effects $\psi_j, j=1,2,3,4$ for $p=450$ and $n=600$ for Scenario 2. The yellow curve represents the GPLM-BAR with AIC penalty, and the blue curve represents GPLM-BAR method with BIC penalty.}\label{ESTS2p450n600}
\end{figure}

\begin{figure}[H]
\centering
\includegraphics[height=12cm, width=16cm]{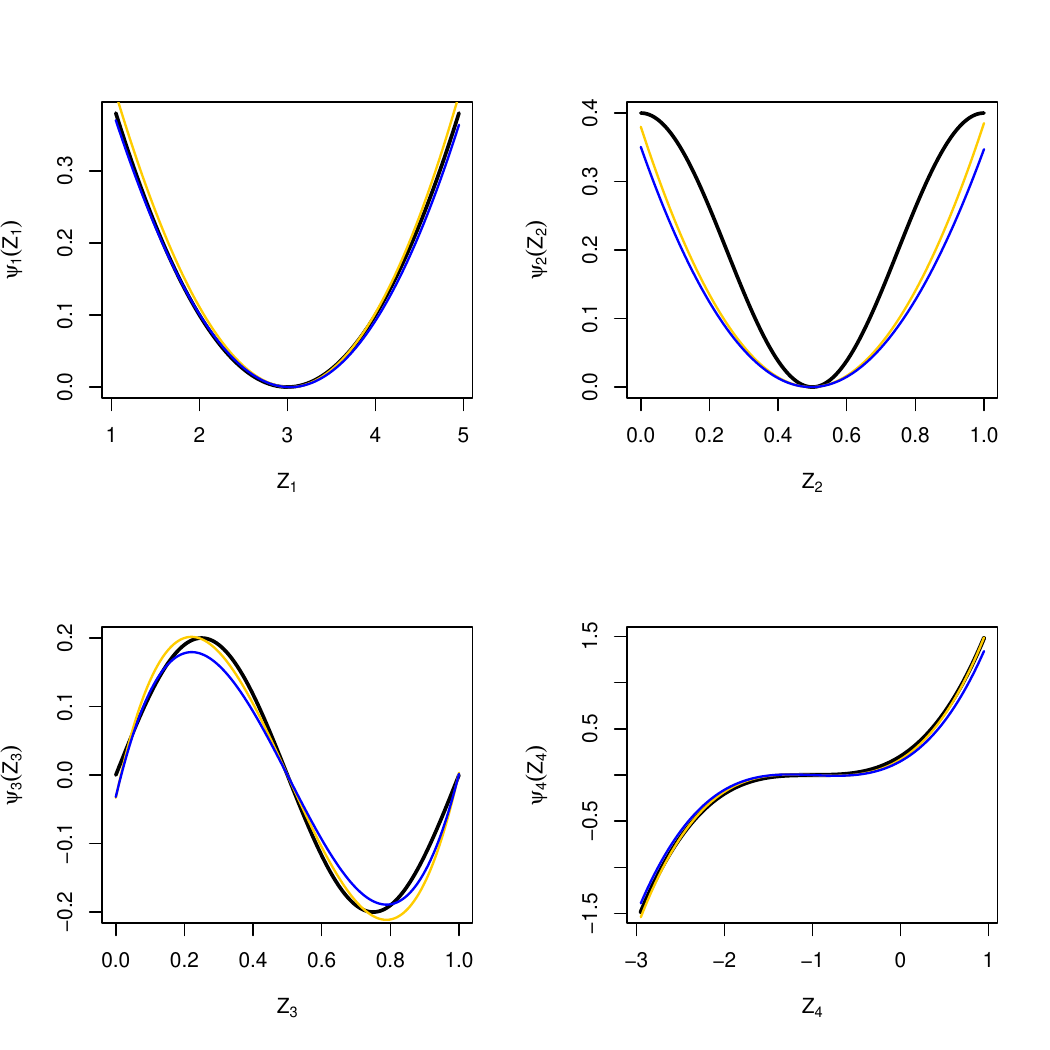}
\caption{Estimated nonlinear covariate effects $\psi_j, j=1,2,3,4$ for $p=450$ and $n=800$ for Scenario 2. The yellow curve represents the GPLM-BAR with AIC penalty, and the blue curve represents GPLM-BAR method with BIC penalty.}\label{ESTS2p450n800}
\end{figure}

\newpage
\section*{More real data analysis results}
\begin{table}[H]
\centering
\caption{SNPs and its associated gene selected by the GPLM-BAR method.} \label{SNPgene}
\begin{tabular}{c|c}
\hline
SNP & Gene \\
\hline
   rs1407961  & ZMYM2 \\
    rs8037353 & GABRG3 \\
    rs7107322 & PRCP \\
    rs2387952 & ADARB2 \\
    rs6680365 & CAMTA1 \\
    rs3136558 & IL1B \\
    rs4131888 & SLC7A11 \\
    rs3845439 & CACNA1E\\
    rs769449 & APOE \\
    rs9549675 & F10 \\
    rs2805543 & ADARB2 \\
    rs821292 & GFOD1 \\
    rs12612481 & PDE11A \\
    rs7188981 & RBFOX1  \\
    rs9932172 & RBFOX1 \\
    rs9282537 & ABCA1 \\
    rs244072 & ADA \\
    rs11859718 & CDH13 \\
    rs17585580 & CNTN5 \\
\hline
\end{tabular}
\end{table}


\end{document}